\documentclass[sigconf,screen]{acmart}
\setcopyright{acmcopyright}
\copyrightyear{2025}
\acmYear{2025}
\acmDOI{XXXXXXX.XXXXXXX}
\acmConference[FSE '25]{International Conference on the Foundations of Software Engineering}{June 23--27,
  2025}{Trondheim, Norway}
\acmBooktitle{Companion Proceedings of the 33rd ACM Symposium on the Foundations of Software Engineering (FSE '25), June 23--27, 2025, Trondheim, Norway}
\acmISBN{}

\usepackage{xspace}

\usepackage{caption,subcaption}
\usepackage[inline]{enumitem}
\usepackage{mathtools}
\usepackage{makecell}
\usepackage{cleveref,hyperref}
\usepackage{textcomp}
\usepackage{booktabs}
\usepackage{multirow}
\usepackage{pgfplots}
\usepackage{pgfplotstable}
\usepackage{filecontents}

\usepackage{tikz}
\usetikzlibrary{shapes.geometric,backgrounds,calc,positioning,fit}
\usepackage{xcolor}

\definecolor{myBlue}{RGB}{0,133,255}
\definecolor{mylilas}{RGB}{148,87,148}
\definecolor{mygreen}{RGB}{2,75,48}
\definecolor{myBlue}{RGB}{0,133,255}
\definecolor{Brown}{RGB}{150,75,0}
\definecolor{Blue}{RGB}{0,0,255}

\definecolor{myOrange}{rgb}{1,0.5,0}
 
\newcommand\phase[1]{\tikz[baseline=(X.base)]\node [draw=myBlue,fill=myBlue,thick,rectangle,inner sep=2pt, rounded corners=2pt](X){\color{white}\textbf{#1}};}

\newcommand{\simulink}{Simulink\textregistered\xspace}

\newcommand\testsequence{\ensuremath{\texttt{TS}}\xspace}

\usepackage{listings}

\usepackage{cleveref}
\usepackage[normalem]{ulem} % for \sout
\usepackage{amsfonts}
\newboolean{showcomments}
\setboolean{showcomments}{true}
\ifthenelse{\boolean{showcomments}}
{\newcommand{\nb}[2]{
  \fcolorbox{black}{yellow}{\bfseries\sffamily\scriptsize#1}
  {\sf\small$\blacktriangleright$\textit{#2}$\blacktriangleleft$}
 }
 
}
{\newcommand{\nb}[2]{}
 
}

\usepackage{colortbl}
\usepackage{amsmath,amsfonts}
\usepackage{algorithmic}
\usepackage{textcomp}

\definecolor{keywordcolor}{RGB}{127,0,85}

\definecolor{background}{rgb}{0.94,0.95,0.96}

\definecolor{preconditionColor}{HTML}{e6ffe0}
\definecolor{commentColor}{HTML}{006e00}
\definecolor{tableColor}{HTML}{fcf8e8}
\definecolor{secondsColor}{HTML}{aa14d2}

\definecolor{preconditionColor}{HTML}{e6ffe0}

\definecolor{postconditionColor}{HTML}{ffffcc}

\usepackage{cleveref}

\newcommand{\speedHecate}{\texttt{Speed\_Hecate\_1}\xspace}
\newcommand{\stepThree}{\texttt{step\_3}\xspace}
\newcommand{\speedHecateOne}{\texttt{Speed\_Hecate\_2}\xspace}
\newcommand{\stepOne}{\texttt{step\_1}\xspace}
\newcommand{\stepTwo}{\texttt{step\_2}\xspace}
\newcommand{\stepFour}{\texttt{step\_4}\xspace}

\newcommand{\stepSeven}{\texttt{step\_7}\xspace}

\newcommand{\hecateParam}[1]{{\textbf{\texttt{#1}}}\xspace}
\newcommand{\inputValue}[1]{{\textbf{\texttt{#1}}}\xspace}
\newcommand{\blue}[1]{\textcolor{Blue}{\textbf{\texttt{#1}}}\xspace}
\newcommand{\step}[1]{\textcolor{Brown}{\textbf{\texttt{#1}}}\xspace}

\newcommand{\timeCondition}[1]{(#1, \textcolor{secondsColor}{'sec'})\xspace}

\newcommand{\system}{\ensuremath{S}\xspace}
\newcommand{\fitness}{\ensuremath{f}\xspace}

\newcommand{\budget}{\ensuremath{T}\xspace}

\newcommand{\testassessment}{\ensuremath{TA}}

\newcommand{\averageTime}{\ensuremath{1\,h\,17\,min\,26\,s}\xspace}
\newcommand{\minTime}{\ensuremath{11\,min\,56\,s}\xspace}
\newcommand{\maxTime}{\ensuremath{8\,h\,16\,min\,22\,s}\xspace}
\newcommand{\stdTime}{\ensuremath{1\,h\,50\,min\,34\,s}\xspace}

\newcommand{\numExperiments}{\ensuremath{36}\xspace}
\newcommand{\numExperimentsFalsified}{\ensuremath{30}\xspace}
\newcommand{\percentageExperimentsFalsified}{\ensuremath{\approx83}\%\xspace}
\usepackage[utf8]{inputenc}

\usepackage[framemethod=TikZ]{mdframed}
\mdfsetup{nobreak=false}
\newenvironment{Answer}[1][]{%
  \ifstrempty{#1}%
  {\mdfsetup{%
    frametitle={%
      \tikz[baseline=(current bounding box.east),outer sep=0pt]
      \node[line width=0pt,anchor=east,rectangle,draw=white,fill=white]
    ;}}
  }%
  {\mdfsetup{%
    frametitle={%
      \tikz[baseline=(current bounding box.east),outer sep=0pt]
      \node[anchor=east,rectangle,draw=white,fill=white]
    {\strut #1};}}%
  }%
  \mdfsetup{innertopmargin=-5pt,linecolor=black,%
            linewidth=0.5pt,topline=true,%
            frametitleaboveskip=\dimexpr-\ht\strutbox\relax,skipabove=\topskip,skipbelow=\topskip}
  \begin{mdframed}[nobreak=false]\relax
  }{\end{mdframed}}
  
\begin{document}

\tikzset{
  basic box/.style = {
    shape = rectangle,
    align = center,
    draw  = #1,
    fill  = #1!25,
    rounded corners},
  header node/.style = {
    font          = \strut\Large\ttfamily,
    text depth    = +0pt,
    fill          = white,
    draw},
  header/.style = {%
    inner ysep = +1.5em,
    append after command = {
      \pgfextra{\let\TikZlastnode\tikzlastnode}
      node [header node] (header-\TikZlastnode) at (\TikZlastnode.north) {#1}
    }
  },
  hv/.style = {to path = {-|(\tikztotarget)\tikztonodes}},
  vh/.style = {to path = {|-(\tikztotarget)\tikztonodes}},
  fat blue line/.style = {ultra thick, blue}
}

%
%\linenumbers

\title{Test Case Generation for \simulink Models:\\ An Experience from the E-Bike Domain}





%\thanks{Supported by organization x.}}

\author[]{Michael Marzella}
\email{m.marzella@studenti.unibg.it}
\orcid{XXXX}
\affiliation{%
  \institution{University of Bergamo}
  %\streetaddress{Viale Marconi 5}
  \city{Dalmine}
  \state{BG}
  \country{Italy}
  %\postcode{24044}
}

\author[]{Andrea Bombarda}
\email{andrea.bombarda@unibg.it}
\orcid{0000-0003-4244-9319}
\affiliation{%
  \institution{University of Bergamo}
  %\streetaddress{Viale Marconi 5}
  \city{Dalmine}
  \state{BG}
  \country{Italy}
  %\postcode{24044}
}

\author[]{Marcello Minervini}
\email{marcello.minervini@unibg.it}
\orcid{0000-0003-4245-445X}
\affiliation{%
  \institution{University of Bergamo}
  %\streetaddress{Viale Marconi 5}
  \city{Dalmine}
  \state{BG}
  \country{Italy}
  %\postcode{24044}
}

\author[]{Nunzio Marco Bisceglia}
\email{n.bisceglia1@studenti.unibg.it}
\orcid{XXXX}
\affiliation{%
  \institution{University of Bergamo}
  %\streetaddress{Viale Marconi 5}
  \city{Dalmine}
  \state{BG}
  \country{Italy}
  %\postcode{24044}
}

\author[]{Angelo Gargantini}
\email{angelo.gargantini@unibg.it}
\orcid{0000-0002-4035-0131}
\affiliation{%
  \institution{University of Bergamo}
  %\streetaddress{Viale Marconi 5}
  \city{Dalmine}
  \state{BG}
  \country{Italy}
  %\postcode{24044}
}

\author[]{Claudio Menghi}
\email{claudio.menghi@unibg.it}
\orcid{0000-0001-5303-8481}
\affiliation{%
  \institution{University of Bergamo}
  %\streetaddress{Viale Marconi 5}
  \city{Dalmine}
  \state{BG}
  \country{Italy}
  %\postcode{24044}
}
\affiliation{%
  \institution{McMaster University}
  %\streetaddress{1280 Main St W}
  \city{Hamilton}
  \state{ON}
  \country{Canada}
  %\postcode{L8S 4L8}
}

\begin{abstract}

Cyber-physical systems development often requires engineers to search for defects in their Simulink models.
Search-based software testing (SBST) is a standard technology that supports this activity.
To increase practical adaption, industries need empirical evidence of the effectiveness and efficiency of (existing) SBST techniques on benchmarks from different domains and of varying complexity.
To address this industrial need, this paper presents our experience assessing the effectiveness and efficiency of SBST in generating failure-revealing test cases for cyber-physical systems requirements.  
Our study subject is within the electric bike (e-Bike) domain and concerns the software controller of an e-Bike motor, particularly its functional, regulatory, and safety requirements.
We assessed the effectiveness and efficiency of HECATE, an SBST framework for Simulink models, to analyze two software controllers. 
HECATE successfully identified failure-revealing test cases for \percentageExperimentsFalsified (\numExperimentsFalsified out of \numExperiments) of our experiments.
It required, on average, \averageTime (\textit{min}=\minTime, \textit{max}=\maxTime, \textit{std}=\stdTime) to compute the failure-revealing test cases.
The developer of the e-Bike model confirmed the failures identified by HECATE.
We present the lessons learned and discuss the relevance of our results for industrial applications, the state of practice improvement, and the results' generalizability.

%the motor controller, while ride-quality requirements address the rider’s perception of smoothness and control, such as torque regulation and stability maintenance under diverse conditions.
%compliance with both  

%
%In this work, we specifically use HECATE to compare two different implementations of the motor controller and identify potentially critical scenarios.
% 

%\red{TODO: Maybe, when we have some numerical data we can add some additional explanations}
\end{abstract}

\maketitle     
\renewcommand{\shortauthors}{Marzella et al.}

\section{Introduction}
\label{sec:introduction}

\emph{Simulink}~\cite{Simulink} is a modeling and simulation language widely used by the cyber-physical system (CPS) industry~\cite{holtmann2024processes,liebel2018model}. It is used by more than 60\% of engineers for  CPS design~\cite{baresi2017test,zheng2015perceptions} and in many domains, such as automotive~\cite{matinnejad2016automated,zander2017model}. 
Simulink appeals to engineers, due to its graphical language suitable for specifying complex systems.
It enables engineers to model their systems by specifying their components and connections~\cite{Simulink}. 
It also offers a large set of add-ons with many pre-designed components tailored for solving problems in different domains~\cite{SimulinkAddOns}. 

\emph{Search-based software testing} (SBST) is widely applied during the development of CPSs to check for software defects~\cite{Arrieta2017,Humeniuk2022}.
It is used in many domains, including real-time, concurrent, distributed, embedded, and safety-critical systems~\cite{Shaukat2010}. 
SBST tools automatically generate test cases to check for violations of system requirements~\cite{Harman2001}, such as safety, functional, and non-functional requirements.

To increase the \emph{industrial applicability} of SBST, it is paramount to empirically evaluate its efficiency and effectiveness and provide practitioners with guidelines and lessons learned that can help them choose the appropriate tools and assess their level of maturity~\cite{Shaukat2010}.
It is also necessary to assess whether SBST techniques scale to realistic development artifacts~\cite{Shaukat2010}.
Indeed, despite being widely recognized as useful tools, SBST test generators' effectiveness and applicability strongly depend on the specific application domain~\cite{harman2015achievements}. Different domains may require different properties from the SBST frameworks.
The research and industrial communities widely recognize the need for replicable experiments and empirical data assessing the benefits of software engineering approaches in practice~\cite{Shaukat2010,dyba2005evidence,kitchenham2004evidence,panichella2018large,sayyad2013parameter,melo2019empirical,7107470,8718592}. 
The need for replicating experiments is of particular importance for Simulink models~\cite{boll2021characteristics,shrestha2023replicability,boll2020replicability,boll2024replicability}, as Simulink projects and models are typically created and maintained in an industrial context and are usually not publicly available due to confidentiality agreements or license
restrictions~\cite{badreddin2013modeling,ding2014open}. 
Therefore, access to these models is limited, making research results hard (if not impossible)
to replicate~\cite{boll2020replicability}.

Several \emph{SBST tools for Simulink} models are available in the literature (e.g.,~\cite{Menghi2020,formica2023search,Donz2010,Waga2020,Yamagata2021,Ernst2019,Formica2024,Zhang2021,Peltomaki2022,Annapureddy2011,Thibeault2021}). 
Many of the models are compared by the falsification category~\cite{Khandait2024,Menghi2023,ErnstABCDFFG0KM21} of the ARCH competition~\cite{ARCHWEBSITE}, an international competition of verification tools for CPSs.
In this paper, we consider HECATE~\cite{Formica2024}.
Unlike the other existing tools, HECATE generates test cases specified as Test Sequences~\cite{TestSequences},
which are automatically adapted to search for requirement violations, and Test Assessments~\cite{TestAssessmentsDescription}, representing the requirements of interest.
Since HECATE directly works with Test Sequences and Test Assessments, which are built-in components within the Simulink Test Framework~\cite{SimulinkTest}, it does not require engineers to learn new modeling languages or frameworks to specify their test oracles and generate their test cases.
This makes HECATE particularly suitable for industrial applications. 
For example, HECATE has been successfully applied to support the development of a cruise controller for an industrial simulator~\cite{Formica2023}, showing its usefulness in finding failure-revealing test cases.
Nevertheless, HECATE is still primarily an academic tool. 
Replicating the experiments to assess HECATE on a different study can provide insightful results about the effectiveness and efficiency of the tool to industrial practitioners~\cite{juristo2012replication,shepperd2018role}. 
The results of our replication are pivotal for technology transfer activities and support industrial adoption of the proposed solution.

This work focuses on the e-Bikes domain. The global e-Bike drive unit market size was USD 27.15 billion in 2022 and is projected to grow from USD 31.85 billion in 2023 to USD 82.84 billion by 2030~\cite{E-BikeMarket}.
We focus on the (software) controller of the electrical motor of the e-Bike.
The software runs on 100\% of e-Bikes~\cite{SofwareBike} and is often designed in Simulink~\cite{E-Bosch}.
It performs many activities, such as governing the motor’s responsiveness to the rider’s speed demands and regulating the battery management. 
For instance, rapid acceleration can lead to faster battery discharge. 
Additionally, the motor controller could enable power regeneration during braking phases.
Given the critical role of the motor controller in e-Bikes, ensuring its reliability and performance is essential, especially as it directly affects user experience, safety, and battery life. Extensive software testing activity is, thus, necessary to address these requirements, as it enables systematic verification of key functionalities, such as speed responsiveness, battery efficiency, and regenerative braking.
This activity is normally performed by physically testing the electric bikes or their components with different loads and scenarios~\cite{Parastiwi2020, Abagnale2016, Gupta2024, Lefticaru2017}.

In this paper, we assess the effectiveness of SBST with HECATE in generating failure-revealing test cases for a study subject from the e-Bike domain.
We considered a complex model of the e-Bike domain and two controllers based on different technologies: the Buck hardware controller and the PWM software controller.
These controllers must satisfy three requirements (functional, regulatory, and safety). 
We also considered six different testing scenarios obtained from different Parameterized Test Sequences.
Therefore, we conducted 36 experiments (2 models $\times$ 3 requirements $\times$ 6 Parameterized Test Sequences).
For each experiment, we ran HECATE and checked whether it could generate a failure-revealing test case.
Our results show that HECATE could effectively generate failure-revealing test cases for \percentageExperimentsFalsified (\numExperimentsFalsified out of \numExperiments) of our experiments. 
We confirmed the failure we found by discussing with the engineer who developed the models.
HECATE required, on average, \averageTime (\textit{min} = \minTime, \textit{max} = \maxTime, \textit{std} = \stdTime).
We critically analyzed our results: We present the lessons learned and discuss the relevance of our results for industrial applications and their generalizability.

This paper is organized as follows.
\Cref{sec:casestudy} presents our study subject from the e-Bike domain. 
\Cref{sec:sbst} describes HECATE.
\Cref{sec:evaluation} presents our evaluation setup and results.
\Cref{sec:discussion} discusses our results.
\Cref{sec:related} outlines the related work.
\Cref{sec:conclusion} presents our conclusions.

\section{E-Bike Study Subject}
\label{sec:casestudy}

In this section, we describe our e-Bike case study~\cite{Marci} focusing on the controlled system and its requirements (\Cref{sec:system}) and the two controllers (\Cref{sec:model}).

\begin{figure*}
    \centering
    \includegraphics[width=\textwidth]{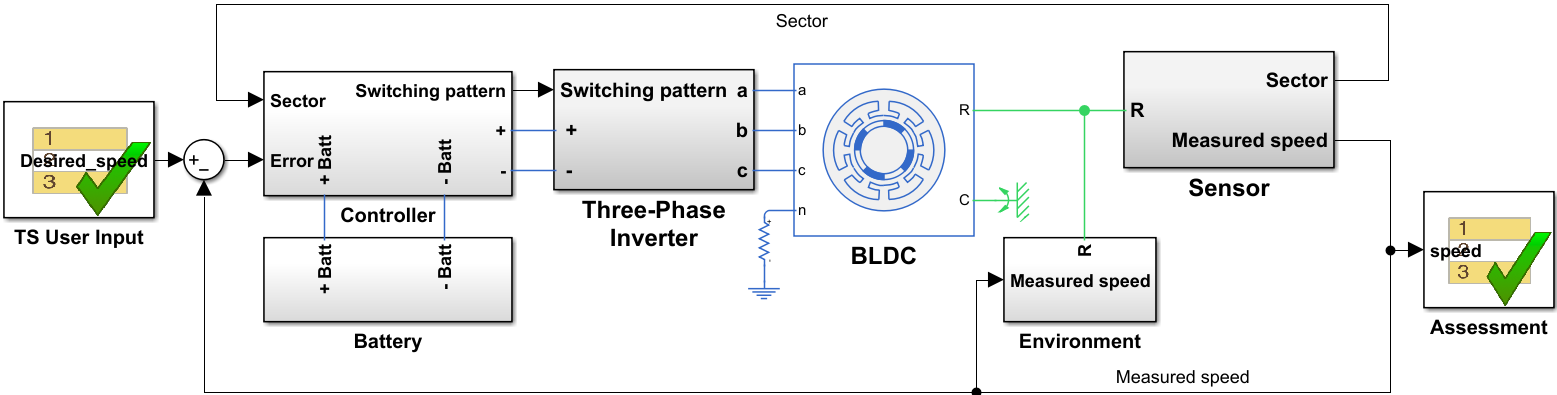}
    \caption{\simulink model for the e-Bike.}
    \Description{\simulink model for the Motor Controller with PWM control}
    \label{fig:pwm}
\end{figure*}

\subsection{The Controlled System} 
\label{sec:system}
\Cref{fig:pwm} presents our study subject from the e-Bike domain.
E-Bikes complement the mechanical power generated by the rider with that provided by an electric motor. Riders can use either a single power source (pedal or battery power alone) or both. The controlled system consists of the following components:
\begin{itemize}
    \item The \emph{User Inputs} component collects the inputs from the rider. 
    The output of the block (\emph{Desired Speed}) models the speed the rider selects over time. The desired speed is used to compute the error (\emph{Error}), i.e., the difference between the \emph{Desired speed} and the \emph{Measured speed} which is one of the inputs of the controller of the e-Bike (\emph{Controller}).
    \item The \emph{Environment} component represents external forces, such as friction and aerodynamic drag. The environment block ensures a realistic simulation of external loads, mimicking the actual resistance an e-Bike would encounter during operation and influencing motor performance. The input is the speed of the e-Bike (\emph{Measured speed}), and the output is a signal that simulates the effects of friction and aerodynamic torque, which is then used to provide feedback to the BLDC Motor through the \emph{R} port connection. 
    \item The \emph{Brushless Direct Current (BLDC) Motor} component converts the electrical energy into rotational motion. A BLDC motor offers higher efficiency and lower maintenance than brushed motors~\cite{BLDC}.
    The inputs of the BLDC are the currents applied to the three phases of the BLDC ($a$, $b$, $c$) and the neutral phase ($n$). 
    The outputs of the BLDC are the torque generated by the motor concerning the rotor (\emph{R}) and the motor case (\emph{C}).
    \item The \emph{Sensor} component monitors the status of the e-Bike by measuring its torque. It returns the active sector (\emph{Sector}) of the BLDC Motor and the e-Bike speed (\emph{Measured speed}).
    \item The \emph{Battery} component is used to store and retrieve electrical energy. A negative current (\emph{- Batt}) recharges the battery. 
    A positive current (\emph{+ Batt}) is used to access the energy stored within the battery.
    \item The \emph{Inverter} component converts the direct current into alternating current and regulates the electrical energy that flows from the battery to the motor and vice versa~\cite{Musumeci2021}.
    When the e-Bike slows down (e.g., during braking), the rotor keeps rotating due to the vehicle's inertia, and the produced energy is used to recharge the battery~\cite{MAMUR2020} (\emph{- Batt}). Otherwise, it flows from the battery to the inverter to help the rider (\emph{+ Batt}).
    The inverter acts on the battery and the motor depending on a direct current (DC) signal (\emph{Switching pattern}) received from the software controller (\emph{Controller}).
    \item The \emph{Controller} component selects the \emph{Switching Pattern} depending on the error difference (\emph{Error}) between the desired speed and the measured speed, and the active sector of the motor (\emph{Sector}). 
\end{itemize}

Engineers design the e-Bike controller (\emph{Controller}) to satisfy the e-Bike requirements from
\Cref{tab:reqs}. 

\begin{itemize}
    \item Requirement R1 is a \emph{functional} requirement: It demands the speed of the motor not to be negative. During braking phases, the motor shall rotate in the same direction while regenerating energy to be stored in the battery: A negative speed is not considered since the e-Bike is assumed to move only in the forward direction. The braking process, therefore, brings the e-Bike from a positive speed to a reduced positive speed (or zero speed). %may lead to unwanted currents and break some of the electrical components in the e-Bike control system.
    \item Requirement R2 is a \emph{regulatory} requirement: It enables electric bikes to assist riders only below 25$km/h$ (i.e., 170 $rpm$ wheel speed, considering a 28 inch wheel), as mandated by most European countries ~\cite{Schepers2018safety,Schleinitz2017}.
\item Requirement R3 is a \emph{safety} requirement: The motor speed shall not exceed the speed requested by the rider.
\end{itemize}

%In this case, indeed, the rider may incur accidents due to the unexpected high speed.

%These requirements have been selected to ensure the motor operates safely and within regulatory limits, maintaining rider control and compliance with legal speed constraints.

\begin{table}[t]
\footnotesize
    \caption{Requirements for the e-Bike.}
    \label{tab:reqs}
    \begin{tabular}{p{.05\linewidth}p{.8\linewidth}}
    \toprule
    \textbf{ID} & \textbf{Description}  \\ \midrule
    R1 & Motor speed shall always be positive or zero. \\
    R2 & Motor speed shall always be lower than 170 rpm. \\
    R3 & Motor speed shall not exceed that requested by the rider. \\\bottomrule
    \end{tabular}
\end{table}

\begin{figure*}
    \centering
     \begin{subfigure}[b]{0.45\textwidth}
    \includegraphics[width=\linewidth]{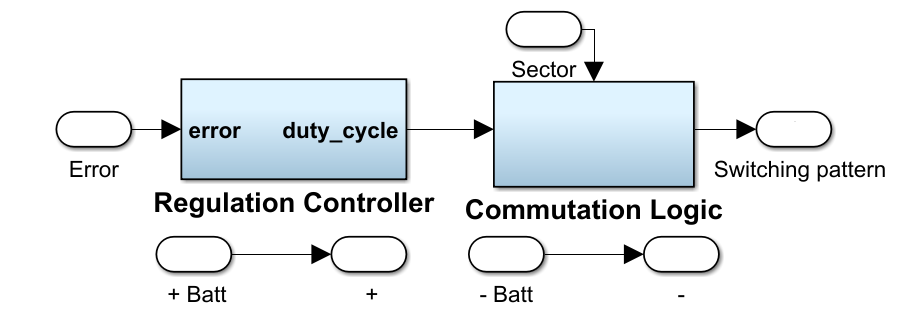}
    \caption{PWM Controller}
    \label{fig:PWM}
    \end{subfigure}
    \hfill
    \begin{subfigure}[b]{0.48\textwidth}
    \includegraphics[width=\linewidth]{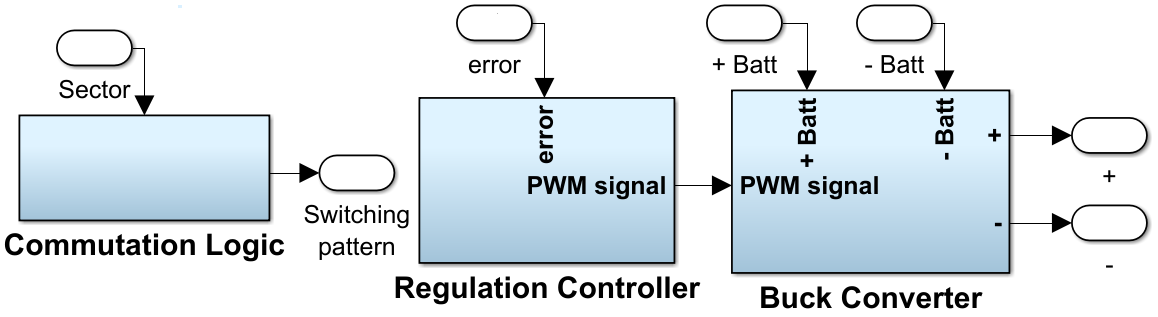}
    \caption{Buck Controller}
    \label{fig:BUCK}
    \end{subfigure}
    \caption{Two software controllers for the e-Bike.}
    \label{fig:controllers}
\end{figure*}
%Engineers are developing models for the controller to ensure the e-Bike satisfies the requirements from \Cref{tab:reqs}.

\subsection{Software Controllers} \label{sec:model}

%\andrea{Dubbio da discutere oggi: vogliamo dire che i modelli sono stati sviluppati per una bicicletta ``vera'' e non sono ``giocattolo'' in mood da giustificare la industry track?}

We considered two controllers (\Cref{fig:controllers}) for the e-Bike: The Pulse Width Modulation (PWM) software controller (\Cref{fig:PWM}) and the Buck hardware controller (\Cref{fig:BUCK}).
Experts from electrical engineering developed these controllers in a project on improving ``green'' mobility solutions, including e-Bikes, and involving several companies, such as Brembo~\cite{Brembo} and Pirelli~\cite{Pirelli}.
Engineers have been developing these models to determine which architecture ensures the highest efficiency.
The development of these models took approximately 100 hours each (including testing activities)~\cite{Corti2024}.

In what follows, we describe the two models:

\begin{itemize}
    \item \emph{PWM (Pulse Width Modulation)} Controller (\Cref{fig:PWM}).
The PWM software controller regulates the \emph{Duty Cycle} of a signal (i.e., the proportion of time the pulse of a signal is active) to modulate the power supplied to the motor. 
Specifically, the controller consists of two subcomponents. A \emph{Regulation Controller} takes the \emph{Error} between the \emph{Desired Speed} and the \emph{Measured Speed} as input and outputs a \emph{Duty Cycle}. That \emph{Duty Cycle} is received as input by the \emph{Commutation Logic}, which also receives \emph{Sector} from the \emph{Sensor}. It outputs the \emph{Switching Pattern}, which determines the sequence of powering the BLDC motor phases for smooth and controlled rotation.  
This controller implements two different PWM algorithms, one for the motoring function and one for the regenerating (braking) function. To switch between one and another the controller uses a binary signal (0 if the desired speed is higher than the effective and 1 when the vice versa is true, i.e., braking signal).
    \item \emph{Buck Controller}  (\Cref{fig:BUCK}).
The Buck hardware controller is divided into subcomponents. 
Unlike the PWM, for the Buck controller, the \emph{Commutation Logic} takes only the \emph{Sector} as input and outputs the \emph{Switching Pattern} for the BLDC motor, managing the phase sequencing. 
%The \emph{Regulation Controller} receives the \emph{Error} as input and outputs the \emph{Duty Cycle} to a \emph{Buck Converter}. Powered by the \emph{Battery}, the \emph{Buck Converter} links the \emph{Inverter} and modulates the power flow between the \emph{Battery} and BLDC. 

\end{itemize}

In this work, the e-Bike engineers provided us with two \emph{intermediate versions} of the PWM and Buck controllers. 
These models do not represent the final deployment-ready models but two intermediate versions that engineers consider relatively stable and ready for the preliminary testing activities.
Typically, engineers extensively test their models and controllers before deployment to check for failures. 
However, since these models are intermediate and still under development, 
before our testing activity started, they were only assessed by considering a limited set of test cases encoding standard profiles for the desired speed.

\section{HECATE}
\label{sec:sbst}

\begin{figure*}
    \begin{subfigure}[b]{0.49\textwidth}
    \centering
    {
    \setlength{\tabcolsep}{0.2em}
    \footnotesize
    \begin{tabular}{l l l}
    \toprule
    \textbf{Step} & \textbf{Transition}   & \textbf{Next Step} \\ \midrule
    \rowcolor{tableColor} \Gape[0pt][2pt]{\makecell[l]{\step{\stepOne} \\ \blue{getSimulationTime()} \\ \inputValue{speed} = \inputValue{  2 *} \blue{getSimulationTime()};}} & \blue{after}\timeCondition{5} & \stepTwo \\
    \makecell[l]{\step{\stepTwo} \\ \inputValue{speed} = \inputValue{10}} & \blue{after}\timeCondition{5} & \stepThree \\ 
    \rowcolor{tableColor} \Gape[0pt][2pt]{\makecell[l]{\step{\stepThree} \\ \inputValue{speed} = \inputValue{10 + 2 * (} \blue{getSimulationTime()} \inputValue{-10)};}}  & \blue{after}\timeCondition{5} & \stepFour \\ 
      \Gape[0pt][2pt]{...} & ... & ..\\

%    \makecell[l]{\step{\stepFour} \\ \inputValue{speed} = \inputValue{0}} & \blue{after}\timeCondition{5} & \stepFive \\
%    \rowcolor{tableColor} \Gape[0pt][2pt]{\makecell[l]{\step{\stepFive} \\ \inputValue{speed} = \inputValue{0 + (Hecate\_sp-0) / (5-0) * (} \blue{getSimulationTime()} \inputValue{-20)}}}  & \blue{after}\timeCondition{5} & \stepSix \\ 
%    \makecell[l]{\step{\stepSix} \\ \inputValue{speed} = \inputValue{Hecate\_sp}} & \blue{after}\timeCondition{5} & \stepSeven \\
    \rowcolor{tableColor} \Gape[0pt][2pt]{\makecell[l]{\step{\stepSeven} \\ \inputValue{speed} = \inputValue{10 + 2 * (} \blue{getSimulationTime()} \inputValue{-30)};}}  & \blue{after}\timeCondition{5} &  \\\bottomrule
    \end{tabular}
    }
    \caption{Test Sequence Block.}
    \label{fig:test_blocks_a}
    \end{subfigure}
    \hfill
    \begin{subfigure}[b]{0.49\textwidth}
    \centering
    {
    \setlength{\tabcolsep}{0.2em}
    \footnotesize
    \begin{tabular}{l l l}
    \toprule
    \textbf{Step} & \textbf{Transition}   & \textbf{Next Step} \\ \midrule
    \rowcolor{tableColor} \Gape[0pt][2pt]{\makecell[l]{\step{\stepOne} \\ }} & \blue{after}\timeCondition{6} & \speedHecate \\
    \makecell[l]{\step{\speedHecate} \\ \blue{verify}(\inputValue{speed} $<=$ 11)} & \blue{after}\timeCondition{4} & \stepThree \\ 
    \rowcolor{tableColor} \Gape[0pt][2pt]{\makecell[l]{\step{\stepThree} \\ }}  & \blue{after}\timeCondition{16} & \speedHecateOne \\ 
    \makecell[l]{\step{\speedHecateOne} \\ \blue{verify}(\inputValue{speed} $<=$ 11)} & \blue{after}\timeCondition{4} & \stepOne \\\bottomrule
    \end{tabular}
    }
    \caption{Test Assessment Block.}
    \label{fig:test_blocks_b}
    \end{subfigure}
    \caption{Test Blocks for our e-Bike model.}
    \label{fig:test_blocks}
\end{figure*}

HECATE is an SBST tool for \simulink models. 
Unlike existing SBST tools, 
HECATE supports \simulink Test Blocks.
Specifically, HECATE identifies failure-revealing test cases represented as Test Sequences using the information from Test Assessments blocks. 
Test Sequence and Test Assessment blocks are embedded within the Simulink model. For example, engineers add a Test Sequence to the model from \Cref{fig:pwm} by replacing the block \emph{User Input} with the Test Sequence block \emph{TS User Input} that generates a \emph{Desired speed} signal used for testing purposes.
They also add the Test Assessment block \emph{Assessment} that receives the measured speed of the vehicle as input. %\claudio{Add these two blocks in the figure}
\Cref{fig:test_blocks} details the Test Sequence (\Cref{fig:test_blocks_a}) and a Test Assessment block (\Cref{fig:test_blocks_b}) from \Cref{fig:pwm}.

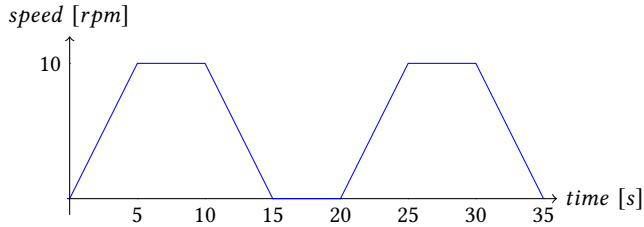
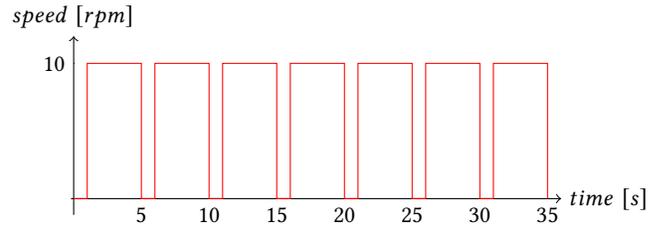
\begin{figure*}
    \begin{subfigure}[b]{0.48\textwidth}
        \centering
        \begin{tikzpicture}[scale=0.18, domain=0:10]
    
          \draw[->] (-0.2,0) -- (36,0) node[right] {$time\;[s]$};
          \draw[->] (0,-1.2) -- (0,12) node[above] {$speed\;[rpm]$};
          \draw[] (0,10) -- (0,10) node[left] {$10$};
          \draw[] (5,0) -- (5,0) node[below] {$5$};
          \draw[] (10,0) -- (10,0) node[below] {$10$};
          \draw[] (15,0) -- (15,0) node[below] {$15$};
          \draw[] (20,0) -- (20,0) node[below] {$20$};
          \draw[] (25,0) -- (25,0) node[below] {$25$};
          \draw[] (30,0) -- (30,0) node[below] {$30$};
          \draw[] (35,0) -- (35,0) node[below] {$35$};
    
          \draw[color=blue] plot coordinates {(0,0) (5,10) (10,10) (15,0) (20,0) (25,10) (30,10) (35,0)};
    
        \end{tikzpicture}
        \caption{Truncated pyramid input with \hecateParam{Hecate\_sp} equal to 10.}
        \label{fig:signal_types_b}
    \end{subfigure}
    \hfill
    \begin{subfigure}[b]{0.48\textwidth}
        \centering
        \begin{tikzpicture}[scale=0.18, domain=0:10]
    
          \draw[->] (-0.2,0) -- (36,0) node[right] {$time \; [s]$};
          \draw[->] (0,-1.2) -- (0,12) node[above] {$speed \; [rpm]$};
          \draw[] (0,10) -- (0,10) node[left] {$10$};
          \draw[] (5,0) -- (5,0) node[below] {$5$};
          \draw[] (10,0) -- (10,0) node[below] {$10$};
          \draw[] (15,0) -- (15,0) node[below] {$15$};
          \draw[] (20,0) -- (20,0) node[below] {$20$};
          \draw[] (25,0) -- (25,0) node[below] {$25$};
          \draw[] (30,0) -- (30,0) node[below] {$30$};
          \draw[] (35,0) -- (35,0) node[below] {$35$};

          \draw[color=red] plot coordinates {(0,0) (1,0) (1,10) (5,10) (5,0) (6,0) (6,10) (10,10) (10, 0) (11,0) (11,10) (15,10) (15,0) (16,0) (16,10) (20,10) (20,0) (21,0) (21,10) (25,10) (25,0) (26,0) (26,10) (30,10) (30,0) (31,0) (31,10) (35,10) (35,0)};
    
        \end{tikzpicture}
        \caption{Rectangular pulse input with \hecateParam{Hecate\_sp} equal to 10.}
        \label{fig:signal_types_a}
    \end{subfigure}
    \caption{Signal types generated as input by HECATE for our e-Bike models.}
    \label{fig:signal_types}
    \Description{Signal types generated as input by HECATE in for our e-Bike models.}
\end{figure*}

A \emph{Test Sequence} defines the test case's input. 
It consists of \emph{steps} connected by \emph{transitions}.
The fragment of the Test Sequence reported in \Cref{fig:test_blocks}, contains four steps (i.e., \step{step\_1}, \step{step\_2}, \step{step\_3}, and \step{step\_7}).
Each test step defines the values to be assumed by the output signals of the Test Sequence. 
For example,  the \step{step\_1} from \Cref{fig:test_blocks} specifies that the value of the \emph{speed} is \emph{10*getSimulationTime()}.
The \step{step\_1} outputs the portion of the \emph{speed} signal from \Cref{fig:signal_types_b} between zero and five seconds.
Transitions define how a Test Block moves across the different steps: They are labeled with a Boolean formula defining the condition for the transition to be fired.
When a transition is fired, the system reaches the step identified by the column \textbf{Next Step}.
For example, the transition from first row of the Test Sequence in \Cref{fig:test_blocks_a} specifies that the test switches from \step{step\_1} to \step{step\_2} after $5\,s$.
When the Simulink model is executed, the Test Sequence generates a signal for its outputs.
For example,  the Test Sequence from \Cref{fig:test_blocks} generates the \emph{speed} output signal from \Cref{fig:signal_types_b}.

A \emph{Test Assessment} block follows the same structure as a Test Sequence block, i.e., it is made by steps and transitions. 
For example, the Test Assessment from \Cref{fig:test_blocks_b} contains four steps (i.e., \step{step\_1} and \step{Speed\_Hecate\_1}, \step{step\_3} and \step{Speed\_Hecate\_2}).
However, unlike Test Sequences, Test Assessment blocks allow engineers to use the \blue{verify} construct. 
This construct verifies whether certain conditions are met when the Test Assessment is in the corresponding test step.
For example, when the Test Assessment is within the test step \step{Speed\_Hecate\_1},  the Test Assessment Block verifies whether the condition \inputValue{speed} $\leq$ 11 holds.

A \emph{test case} consists of a Test Sequence block and a Test Assessment block. 
The Test Sequence block creates input signals supplied to the \simulink simulator. 
The model is then ran with these inputs to simulate the corresponding Test Sequence. 
The Test Assessment block monitors the model’s output signals to determine whether any of its \blue{verify} expressions are violated. 
Typically, a Test Assessment is associated with one or more system requirements.
For example, the Test Assessment in \Cref{fig:test_blocks_b} has been written to check the requirement R3 discussed in \Cref{sec:system}.
Engineers can inspect the satisfaction of these conditions using an appropriate GUI~\cite{TestAssessmentGUI}.  
If (at least) one of the conditions of the Test Assessment is violated, the test case represented by the Test Sequence is failure-revealing, meaning that it violates the conditions of the Test Assessment. 
%Engineers need to inspect their models to identify the cause of the failure.

HECATE~\cite{Formica2024} extends this existing testing framework by supporting SBST.
It enables the automatic generation of Test Sequences driven by a fitness function generated from the Test Assessment block.
Specifically, HECATE requires engineers to extend their Test Sequences into Parameterized Test Sequences.

\begin{figure*}
\begin{subfigure}[b]{0.49\textwidth}    
    \centering
    {
    \setlength{\tabcolsep}{0.2em}
    \footnotesize
   \begin{tabular}{l l l}
   \toprule
    \textbf{Step} & \textbf{Transition}   & \textbf{Next Step} \\ \midrule
    \rowcolor{tableColor} \Gape[0pt][2pt]{\makecell[l]{\step{\stepOne} \\ \blue{getSimTime()} \\ \inputValue{speed}=\inputValue{Hecate\_sp/5*}\blue{getSimTime()};}} & \blue{after}\timeCondition{5} & \stepTwo \\
   \makecell[l]{\step{\stepTwo} \\ \inputValue{speed}=\inputValue{Hecate\_sp}} & \blue{after}\timeCondition{5} & \stepThree \\ 
    \rowcolor{tableColor} \Gape[0pt][2pt]{\makecell[l]{\step{\stepThree} \\ \inputValue{speed}=\inputValue{Hecate\_sp-Hecate\_sp/5*(}\blue{getSimTime()}\inputValue{-10)};}}  & \blue{after}\timeCondition{5} & \stepFour \\ 
     \Gape[0pt][2pt]{...} & ... & ..\\
    \rowcolor{tableColor} \Gape[0pt][2pt]{\makecell[l]{\step{\stepSeven} \\ \inputValue{speed}=\inputValue{Hecate\_sp-Hecate\_sp/5*(}\blue{getSimTime()}\inputValue{-30)};}}  & \blue{after}\timeCondition{5} &  \\
    \bottomrule
    \end{tabular}
     $^\ast$ \blue{getSimulationTime} was shortened into \blue{getSimTime()}.
    }
    \caption{Parameterized Test Sequence Block A.}
    \label{fig:parameterizedTestSequence_a}
    \end{subfigure}
    \hfill
    \begin{subfigure}[b]{0.49\textwidth}    
    \centering
    {
    \setlength{\tabcolsep}{0.2em}
    \footnotesize
   \begin{tabular}{l l l}
   \toprule
    \textbf{Step} & \textbf{Transition}   & \textbf{Next Step} \\ \midrule
    \rowcolor{tableColor} \Gape[0pt][2pt]{\makecell[l]{\step{\stepOne} \\ \blue{getSimTime()} \\ \inputValue{speed}=\inputValue{0};}} & \blue{after}\timeCondition{1} & \stepTwo \\
   \makecell[l]{\step{\stepTwo} \\ \inputValue{speed}=\inputValue{Hecate\_sp}} & \blue{after}\timeCondition{4} & \stepOne \\
    \bottomrule
    \end{tabular}\\
    $^\ast$ \blue{getSimulationTime} was shortened into \blue{getSimTime()}.
    }

    \caption{Parameterized Test Sequence Block B.}
    \label{fig:parameterizedTestSequence_b}
    \end{subfigure}
    \caption{Parameterized Test Sequences.}
    \label{fig:parameterizedTestSequence}
   \end{figure*}
   
A \emph{Parameterized Test Sequence} is a Test Sequence in which some values are replaced by parameters that can be assigned to values produced by HECATE. \Cref{fig:parameterizedTestSequence} shows an example of a Parameterized Test Sequence. 
HECATE can assign different values to the search parameter \inputValue{Hecate\_sp} to generate many test cases from different Test Sequences.
For example, the Test Sequence block from \Cref{fig:test_blocks_a} is an example instance generated from the Parameterized Test Sequence from  \Cref{fig:parameterizedTestSequence} and obtained by assigning the value $20$ to the search parameter \inputValue{Hecate\_sp}.
To generate realistic Test Sequences, engineers can specify lower and upper bounds for the search parameters. For example, engineers can specify that the lower and the upper bound for the values assigned to the parameter \inputValue{Hecate\_sp} should be 0 and 170 rpm.

\tikzstyle{output} = [coordinate]

\begin{figure}[t]
\begin{tikzpicture}[auto,
 block/.style ={rectangle, draw=black, thick, fill=white!20, text width=5em,align=center, rounded corners},
 block1/.style ={rectangle, draw=blue, thick, fill=blue!20, text width=5em,align=center, rounded corners, minimum height=2em},
 line/.style ={draw, thick, -latex',shorten >=2pt},
 cloud/.style ={draw=red, thick, ellipse,fill=red!20,
 minimum height=1em}]
 
\draw (0,0) node[block] (Input) {\phase{1} \footnotesize Test Sequence\\Generation};
\node[block, right of=Input,node distance=2.5cm] (Model){\phase{2} \footnotesize System\\  Execution};
 
\node [block, right of=Model,node distance=2.7cm] (Requirement) {\phase{3} \footnotesize Fitness\\ Assessment};

\node [output, right of=Requirement,node distance=2.3cm] (uoutput) {};
\node [output, above of=Input,node distance=1cm] (Constraints) {};
\node [output, above of=Model,node distance=1cm] (InputModel) {};
\node [output, below of=Requirement,node distance=0.7cm] (v1a) {};
\node [output, left of=v1a,node distance=3cm] (v1) {};
\node [output, below of=Input,node distance=0.7cm] (v2) {};
 \node [output, below of=Requirement,node distance=1.2cm] (mftwo) {};
\node [output, below of=Requirement,node distance=1.2cm] (mf) {};
\node [output, above of=Requirement,node distance=1cm] (ff) {};
\node [output, above of=Requirement,node distance=1cm] (budget1) {};
\node [output, above of=Requirement,node distance=1cm] (budget2) {};
\node [output, left of=budget2,node distance=0.7cm] (budget3) {};

\draw[-stealth] (budget3) -- (Requirement)
    node[midway,above]{$\budget$};

%arrows
\draw[-stealth] (Input.east) -- (Model.west)
    node[midway,above]{$\testsequence$};
%\draw[-stealth] (Constraints.south) -- (Input.north)
%    node[midway,right]{$\asmpt$};
\draw[-stealth] (InputModel.south) -- (Model.north)
    node[midway,right]{$\system$};

\draw[-] (v1) |- (v2)node[midway,below]{$\fitness(\testassessment)$};
\draw[-stealth] (v2) -- (Input.south);
 \draw[-stealth] (Model) -- (Requirement.west) node[midway,above]{$\system(\testsequence)$};
 \draw[-stealth] (Requirement.east) -- (uoutput) node[midway,above]{$\testsequence$/NFF};
 \draw[-stealth] (ff) -- (Requirement) node[midway,right]{$\testassessment$};
 
   \draw[-] (Requirement) -- (v1a) node[midway,left]{};
 
  \draw[-] (v1a) |- (v1) node[midway,left]{};
 \end{tikzpicture}
\caption{Overview of the HECATE framework.}
\label{fig:SBST}
\end{figure}
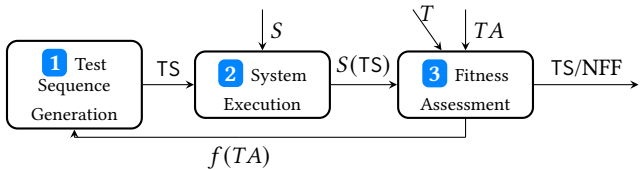

Similar to existing SBST frameworks, HECATE iteratively performs the steps from \Cref{fig:SBST}:
\begin{enumerate}
    \item \emph{Test Sequence Generation}: Generates a new Test Sequence (\emph{TS}) for the system $S$ by assigning values to the parameters in the Parameterized Test Sequence;
    \item \emph{System Execution}: The system model $S$ is executed by providing the input generated from the Test Sequence (\emph{TS}) and generates the output $S(\testsequence)$;
    \item \emph{Fitness Assessment}: A fitness function obtained from the Test Assessment TA is evaluated w.r.t. the output $S(\testsequence)$. HECATE assesses whether the fitness value is below the desired threshold level.
    A Test Sequence is failure-revealing when the computed fitness is smaller than the threshold.
\end{enumerate}

HECATE stops the search procedure if a failure-revealing Test Sequence is found or the maximum time $T$ allotted for the search is reached.

The main advantage of HECATE over the existing framework is that it relies on Test Sequences and Test Assessments, which are part of  \simulink.
Therefore, engineers can start from manually defined test cases, parameterize them, and use them within HECATE.

%We discuss how we use HECATE for our experimentation with our model from the e-Bike domain in the following.
%During the experiments, the standard speed input graph was replaced with a Test Sequence Block comprising six different scenarios: three double-truncated pyramid scenarios and three rectangular pulse scenarios.

For example, in this study, the double truncated pyramid scenario (\Cref{fig:signal_types_b}) consisted of seven segments with a total duration of $35\,s$.
It corresponds to the Parameterized Test Sequence  (\emph{t-pyramid}) detailed in \Cref{fig:parameterizedTestSequence_a}.
Specifically, the 2nd, 4th, and 6th segments feature constant speeds (e.g., see \step{step\_2} in \Cref{fig:parameterizedTestSequence_a}) set to the \hecateParam{Hecate\_sp} generated by the tool.
The 1st and 5th segments (e.g., \step{step\_1} in \Cref{fig:parameterizedTestSequence_a}) are characterized by constant acceleration, while the 3rd and 7th segments (\step{step\_3} and \step{step\_7} in \Cref{fig:parameterizedTestSequence_a}) involve constant deceleration.
The acceleration and deceleration rates are consistent across these segments but can be swapped depending on whether the initial and the 4th segment speeds are lower or higher than the \hecateParam{Hecate\_sp}.
The Test Sequence in \Cref{fig:test_blocks_a} is obtained from the Parameterized Test Sequence from \Cref{fig:parameterizedTestSequence_a} by using $10$ as the value for the \hecateParam{Hecate\_sp} parameter.

Each rectangular pulse scenario (\Cref{fig:signal_types_a}) is characterized by the Parameterized Test Sequence (\emph{rect-pulse}) detailed in \Cref{fig:parameterizedTestSequence_b}. It generates a constant speed phase from $t=0\,s$ to $t=1\,s$ (\step{step\_1} in \Cref{fig:parameterizedTestSequence_b}) with a fixed value.
From $t=1\,s$ to $t=5\,s$ (\step{step\_2} in \Cref{fig:parameterizedTestSequence_b}), a fixed speed, identified by the placeholder \hecateParam{Hecate\_sp}, is generated by the tool.
These two steps are cyclically repeated (see the next step for \step{step\_2} in \Cref{fig:parameterizedTestSequence_b}) for $35\,s$.

\section{Evaluation}
\label{sec:evaluation}

We consider the following research questions:
\begin{itemize}
    \item[\textbf{RQ1}:] How \emph{effective} is HECATE in generating failure-revealing test cases for our e-Bike model? 
    \item[\textbf{RQ2}:] How \emph{efficient} is HECATE in generating failure-revealing test cases for our e-Bike model? 
\end{itemize}

\begin{table}[t]
\small
    \caption{Values assigned to the configuration parameters of the search algorithms used by HECATE.}
    \label{tab:configuration}
    \begin{tabular}{p{.6\linewidth}p{.3\linewidth}}
    \toprule
    \textbf{Parameter} & \textbf{Value}  \\ \midrule
    Optimization solver & Uniform random\\
    Number of runs & 10\\
    Maximum number of iterations per run & 50\\
    \bottomrule
    \end{tabular}
\end{table}

We present the experimental setup (\Cref{sec:setup}).
We then discuss the results for RQ1 (\Cref{sec:effectiveness}) and RQ2 (\Cref{sec:efficiency}).

\begin{table*}[t]
\caption{Experimental results for the e-Bike case study.}
\label{tab:results}
\renewcommand{\arraystretch}{0.8}
\begin{tabular}{@{} l | l  | ll l |ll  l |ll l@{} }
\toprule
&        & \multicolumn{3}{c|}{\textbf{TA\_R1}} & \multicolumn{3}{c|}{\textbf{TA\_R2}} & \multicolumn{3}{c}{\textbf{TA\_R3}} \\ 
\textbf{Model} &  \textbf{PTS} & \textbf{FR}    & $\mathbf{\bar{S}}$    & $\mathbf{\hat{S}}$   & \textbf{FR}    & $\mathbf{\bar{S}}$    & $\mathbf{\hat{S}}$   & \textbf{FR}    & $\mathbf{\bar{S}}$    & $\mathbf{\hat{S}}$   \\ \midrule
\multirow{6}{*}{\textbf{PWM}} 
&  \emph{t-pyramid-0} & 10/10 & 1.6 & 1.0 & 10/10 & 9.6 & 4.0 &       10/10 %(10/10)
&       1.9 %(1.7)
&        1.5 %(1.5)
\\
&  \emph{t-pyramid-85} & 10/10 & 7.6 & 4.0 & 10/10 & 3.0 & 3.0 &       10/10 %(10/10)
&       1.0 %(6.0)
&        1.0 %(3.5)
\\
&  \emph{t-pyramid-130} &       10/10&       10.6&        8.5&       10/10&       1.1&        1.0&       10/10 %(10/10)
&       1.0 %(2.2)
&        1.0 %(2.0)
\\ 
\cmidrule(l){2-11} 
& \emph{rect-pulse-0} & 10/10 & 8.9  & 6.0 & 10/10 & 8.6 & 8.5 &        10/10 %(7/10)
&       1.6 %(20.7)
&       1.0 %(13.0)
\\
&   \emph{rect-pulse-85} & 10/10 & 14.6 & 12.5 & 10/10 & 3.0 & 2.5 &       10/10 %(10/10)
&       1.0 %(1.8)
&        1.0 %(1.5)
\\
&  \emph{rect-pulse-130} &       10/10&       6.7&        4.0&       10/10&       2.2&        1.5&       10/10 %(10/10)
&       1.0 %(1.2)
&        1.0 %(1.0)
\\ 
\midrule
\multirow{6}{*}{\textbf{Buck}} 
& \emph{t-pyramid-0} &       0/10&    --   &    --    &       4/10&       28.8&        26.0&       10/10&       8.1&        3.0\\
& \emph{t-pyramid-85} &       0/10&   --    &     --   &       5/10&       17.6&        22.0&       10/10&       4.3&        4.0\\
& \emph{t-pyramid-130} &       0/10&   --    &   --     &       10/10&       1.9&        1.0&       10/10&       2.1&        1.5\\ 
\cmidrule(l){2-11}
&  \emph{rect-pulse-0} &       0/10&   --    &    --    &       10/10&       2.9&        2.0&       10/10&       7.3&        7.5\\
& \emph{rect-pulse-85} &       0/10&    --   &    --    &       10/10&       2.4&        2.0&       10/10&       1.6&        1.5\\
& \emph{rect-pulse-130} &       0/10& --      &   --     &       10/10&       2.7&        3.0&       10/10&       1.0&        1.0\\ 
\bottomrule
\end{tabular}
\end{table*}

\subsection{Experimental Setup}
\label{sec:setup}
To assess the effectiveness of HECATE in generating failure-revealing test cases, we performed 36 experiments.
Each experiment was obtained by considering one of the two models, one of the three Test Assessments, and one of the six Parameterized Test Sequences.  
The two models were the \simulink models of the e-Bike obtained by considering the PWM and Buck controllers from \Cref{sec:model}.
The three Test Assessments (TA\_R1, TA\_R2, TA\_R3) encode the requirements (functional, regulatory, and safety respectively) from \Cref{sec:system}.
The six Parameterized Test Sequence (\emph{t-pyramid-0}, \emph{t-pyramid-85}, \emph{t-pyramid-130}, \emph{rect-pulse-0}, \emph{rect-pulse-85}, and \emph{rect-pulse-130}) were obtained by considering three versions of two Parameterized Test Sequences from  \Cref{fig:parameterizedTestSequence} (\emph{t-pyramid} and \emph{rect-pulse}) each.
These versions ensured the value of the speed of the rect-pulse and t-pyramid signals is increased by $0$, $85$, and $130$ in each time instant.
Each experiment was obtained by selecting one of two models, one of three requirements, and one of six Parameterized Test sequences.  
Therefore, we ran 36 experiments (2 models $\times$ 3 Test Assessment blocks $\times$ 6 Parameterized Test Sequences) in total.

Each experiment was performed by using the configuration parameters listed in \Cref{tab:configuration}.
We used the Uniform Random solver, which performed uniform sampling in the parameter space, and we
ran HECATE for each experiment by setting the maximum number of search iterations to 10. Every run was repeated 50 times to account for the stochastic nature of the algorithm.
We executed experiments on a consumer-grade laptop with the following specifications: an Intel(R) Core(TM) i7-9750H CPU running at 2.60 GHz, featuring six cores and a 12 MB SmartCache, supported by 16 GB of installed RAM. 
For each experiment, we recorded which of the 10 runs returned a failure-revealing test case.

%The Test Assessment Block was linked to the actual measured speed, allowing for performance evaluation by comparing it with the input from the Test Sequence.
%The Test Assessment included three scenarios designed to validate the requirements outlined in  \Cref{sec:system}, as detailed in \Cref{tab:reqs}. These scenarios assessed the controller's ability to meet specific performance criteria related to speed tracking, stability, and response under varying input conditions.

%Each model underwent testing with all six Test Sequence scenarios and the three Test Assessment scenarios reported in \Cref{tab:results}, resulting in a total of 18 tests per model. 

 \Cref{tab:results} summarizes our results. 
 Each row of the table encodes the model (\textbf{Model}) and the Parameterized Test Sequence (\textbf{PTS}) selected for the experiment.
 The three vertical portions of the table identify the Test Assessment blocks (TA\_R1, TA\_R2, and TA\_R3) considered in our experiments.  
 For each experiment, the table reports the falsification rate (\emph{FR}), i.e., the number of runs in which HECATE could find a failure-revealing test case, the average ($\hat{S}$), and the median ($\bar{S}$) number of iterations required to identify the failure-revealing test case.

\subsection{Effectiveness (RQ1)}
\label{sec:effectiveness}
The results from \Cref{tab:results} for each requirement are as follows.

\textbf{Functional Requirement R1}.
We discuss the results related to the PWM and the Buck controller for the functional requirement R1 (specifying that the speed is not negative) separately.

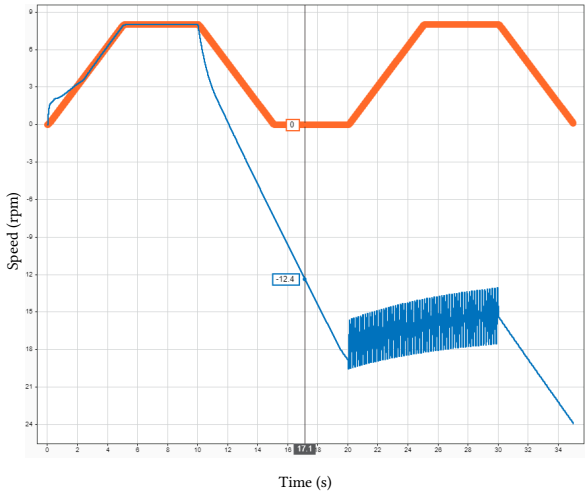
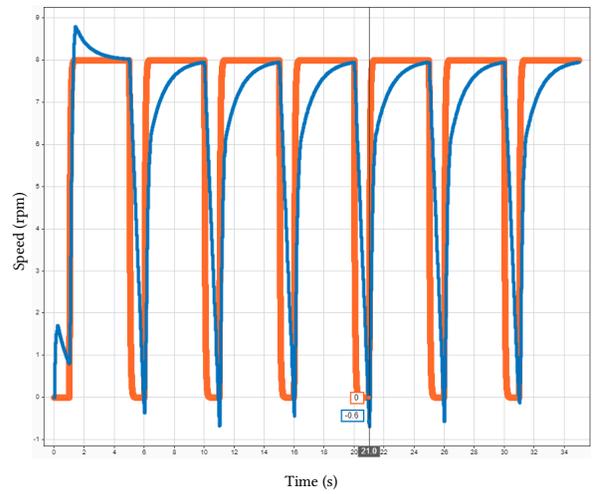
\begin{figure*}[t]
    \centering
     \begin{subfigure}[b]{0.48\textwidth}
     \begin{tikzpicture}
            \node[anchor=south west, inner sep=0] (img) at (0,0) 
            {\includegraphics[width=.87\linewidth]{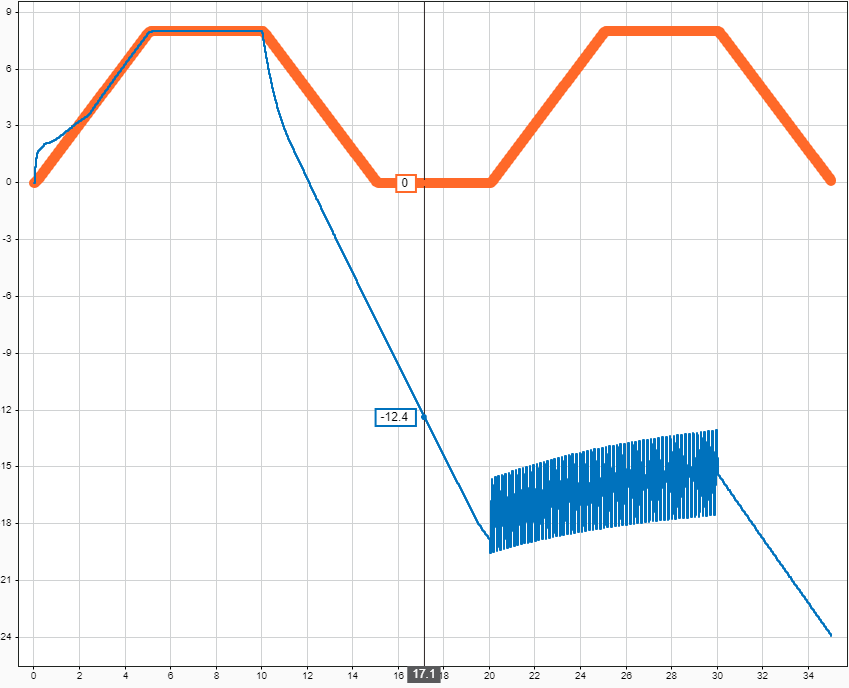}};
            \begin{scope}[x={(img.south east)},y={(img.north west)}]
                % Etichette degli assi
                \node[anchor=north, font=\scriptsize] at (0.5, -0.02) {Time (s)};
                \node[anchor=center, rotate=90, font=\scriptsize] at (-0.02, 0.5) {Speed (rpm)};
            \end{scope}
        \end{tikzpicture} 
    \caption{t-pyramid-0 input}
    \label{fig:Grafico_R1_pyramid}
    \end{subfigure}
    \hfill
    \begin{subfigure}[b]{0.48\textwidth}
    \begin{tikzpicture}
            \node[anchor=south west, inner sep=0] (img) at (0,0) 
            {\includegraphics[width=.87\linewidth]{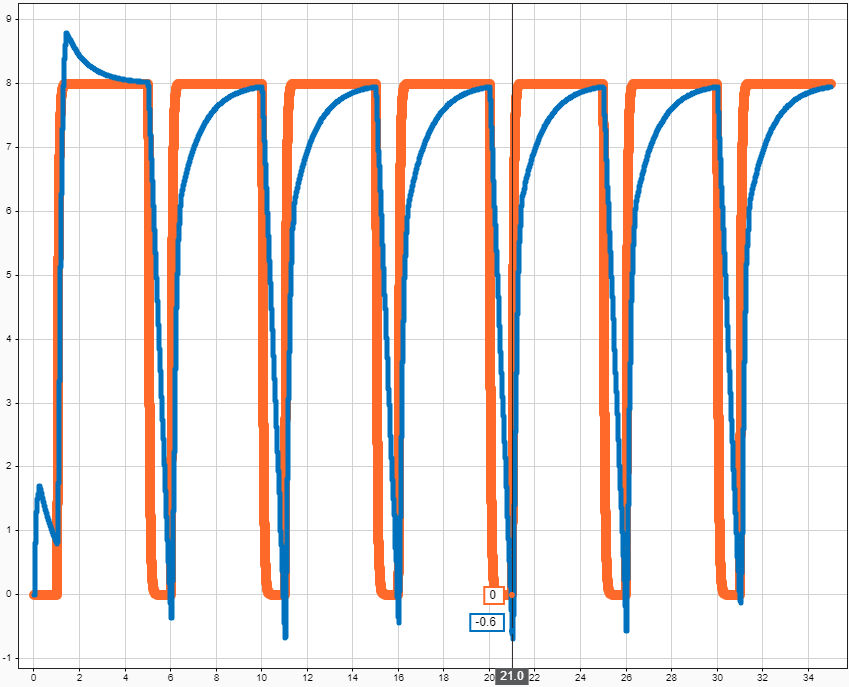}};
            \begin{scope}[x={(img.south east)},y={(img.north west)}]
                % Etichette degli assi
                \node[anchor=north, font=\scriptsize] at (0.5, -0.02) {Time (s)};
                \node[anchor=center, rotate=90, font=\scriptsize] at (-0.02, 0.5) {Speed (rpm)};
            \end{scope}
        \end{tikzpicture}
    \caption{rect-pulse-0 input}
    \label{fig:Grafico_R1_step}
    \end{subfigure}
    \caption{Desired (orange) and actual (blue) speeds of the e-Bike for the PWM  controller and requirement (R1).}
    \label{fig:grafico_r1}
\end{figure*}

\emph{PWM controller}. HECATE could generate a failure-revealing test case for all the experiments related to the PWM controller. 
HECATE showed a 10/10 falsification rate for each experiment for all the considered Test Sequences.  
Therefore, the PWM controller did not ensure that the speed was always greater than or equal to zero. 

The average ($\hat{S}$) and the median ($\bar{S}$) number of iterations required to identify the failure-revealing test case show that for Test Sequences focusing on low-speed values ($0$ and $85$), 
 %that by considering the truncated-pyramid Test and Sequences (\emph{rect-pulse-0}, \emph{rect-pulse-85}, \emph{t-pyramid-0}, \emph{t-pyramid-85}),   
HECATE required more iterations (1.6 vs 8.9 and 7.6 vs 14.6 for the average, and 1.0 vs 6.0 and 4.0 vs 6.0 for the median)  to identify the failure-revealing test cases for the square wave signal (\emph{rect-pulse}) than the truncated-pyramid Test Sequences (\emph{t-pyramid}).
The results show an opposite trend at higher speeds (\emph{rect-pulse-130}, \emph{t-pyramid-130}).

\Cref{fig:grafico_r1} shows the measured speed (blue) and the desired speed (orange) of two failure-revealing test cases generated by the \emph{t-pyramid-0} and  \emph{rect-pulse-0} Parameterized Test Sequences.
Specifically, for the \emph{t-pyramid-0} and \emph{rect-pulse-0} test cases, eight was the highest value of the desired speed (orange), as this was the value selected for the \textbf{Hecate\_sp} parameter. 
The lower bound was 0 since these Parameterized Test Sequences correspond to the one from \Cref{fig:parameterizedTestSequence} (\emph{t-pyramid-0} and \emph{rect-pulse-0}): No increment on the speed values was applied.

%\claudio{@Marcello puoi aggiungere un commento/riflessione personale?}
 
 %\claudio{TODOPaper usare random dice poco. vanno replicate gli esperimenti con SA.}
 %\claudio{TODOPaper per version finale del paper: Chiedere feedback a Marcello.}
%The tests revealed distinct behaviors between the rectangular pulse and scenarios. Rectangular pulse scenarios, characterized by abrupt changes in speed, resulted in one iteration for failure detection, particularly at intermediate speeds (SP = 85), suggesting that the model struggles with sudden changes. In contrast, truncated-pyramid scenarios, with gradual accelerations and decelerations, led to quicker failure detection at lower speeds (SP = 0), but required more iterations at higher speeds (SP = 130). These results indicated that the PWM model responds differently to sudden versus gradual input changes, with a tendency for more complex or slower failure detection in certain conditions.

\emph{Buck Controller.} Unlike the PWM controller, HECATE could not generate any failure-revealing test case for all the experiments of the Buck controller for Requirement R1.

\textbf{Regulatory Requirement R2.}
\label{sec:resR2}
In this segment, we discuss the results for the functional requirement R2.

\emph{PWM controller}. HECATE generated a failure-revealing test case for all the experiments performed on the PWM controller, with a 10/10 falsification rate for each experiment with all the Test Sequences we considered.
Thus, the PWM controller may lead the bike to violate the regulatory requirement, with the e-Bike overpassing the maximum speed (170 rpm) from the regulations.

%The average ($\hat{S}$) and the median ($\bar{S}$) number of iterations required to identify the failure-revealing test case show that in low-speed scenarios ($0$ and $85$), 
%both with truncated-pyramid and rectangular pulse Test Sequences (\emph{rect-pulse-0}, \emph{rect-pulse-85}, \emph{t-pyramid-0}, \emph{t-pyramid-85}) HECATE requires more iterations to identify the failure-revealing test cases, showing better stability at lower speeds.
%This trend might suggest a system with effective control at lower speeds but faces challenges at higher starting speeds. This could be attributed to the dynamics of speed regulation or the response time of the controller. 

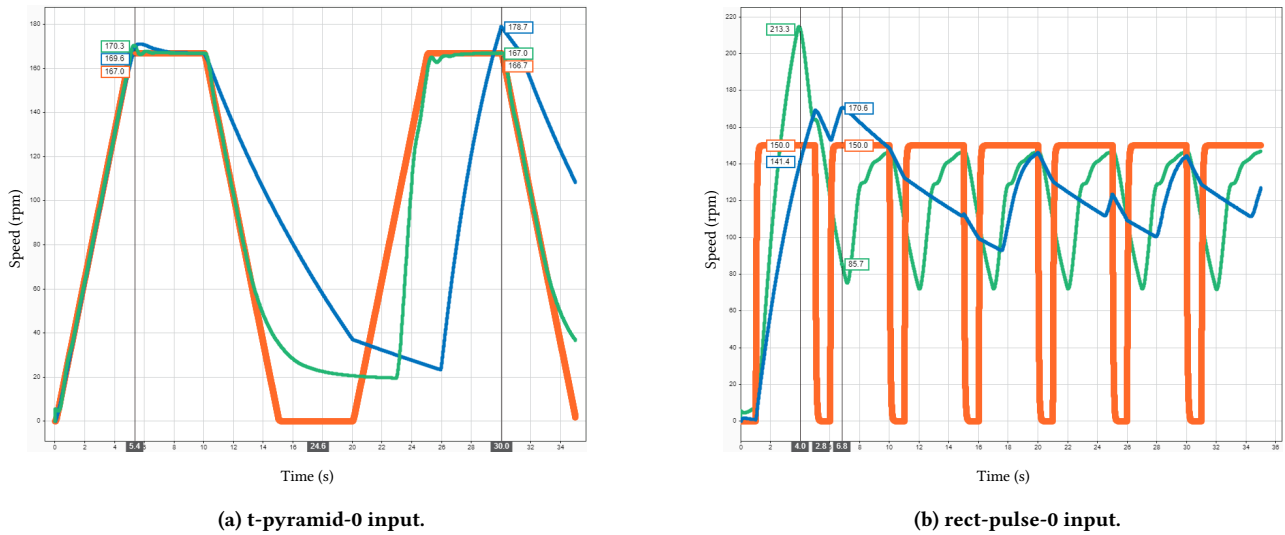
\begin{figure*}
    \centering
     \begin{subfigure}[b]{0.48\textwidth}
    \begin{tikzpicture}
            \node[anchor=south west, inner sep=0] (img) at (0,0) 
            {\includegraphics[width=.87\linewidth]{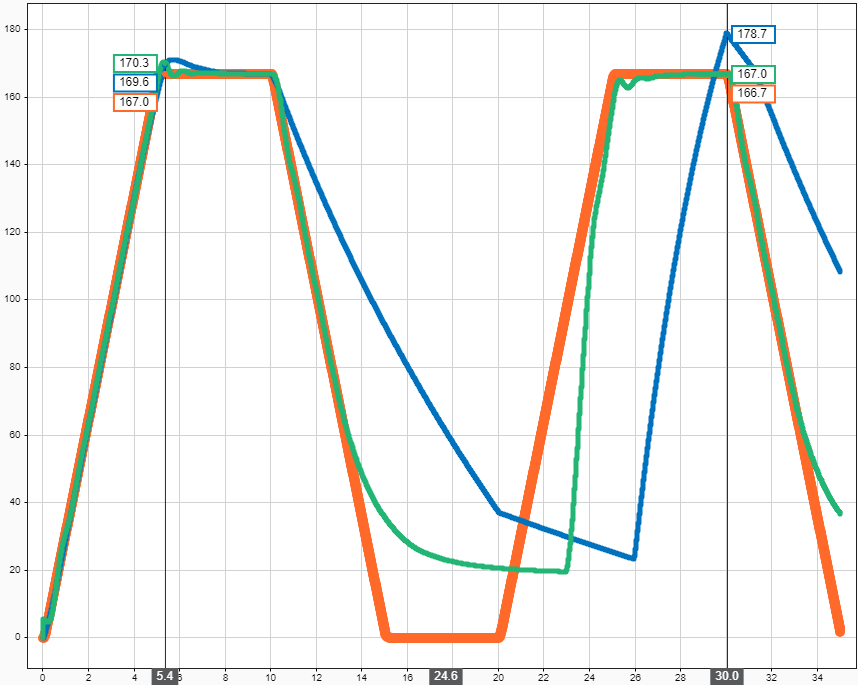}};
            \begin{scope}[x={(img.south east)},y={(img.north west)}]
                % Etichette degli assi
                \node[anchor=north, font=\scriptsize] at (0.5, -0.02) {Time (s)};
                \node[anchor=center, rotate=90, font=\scriptsize] at (-0.02, 0.5) {Speed (rpm)};
            \end{scope}
        \end{tikzpicture}
    \caption{t-pyramid-0 input.}
    \label{fig:R2_graph_pyramid}
    \end{subfigure}
    \hfill
    \begin{subfigure}[b]{0.48\textwidth}
    \begin{tikzpicture}
            \node[anchor=south west, inner sep=0] (img) at (0,0) 
            {\includegraphics[width=.87\linewidth]{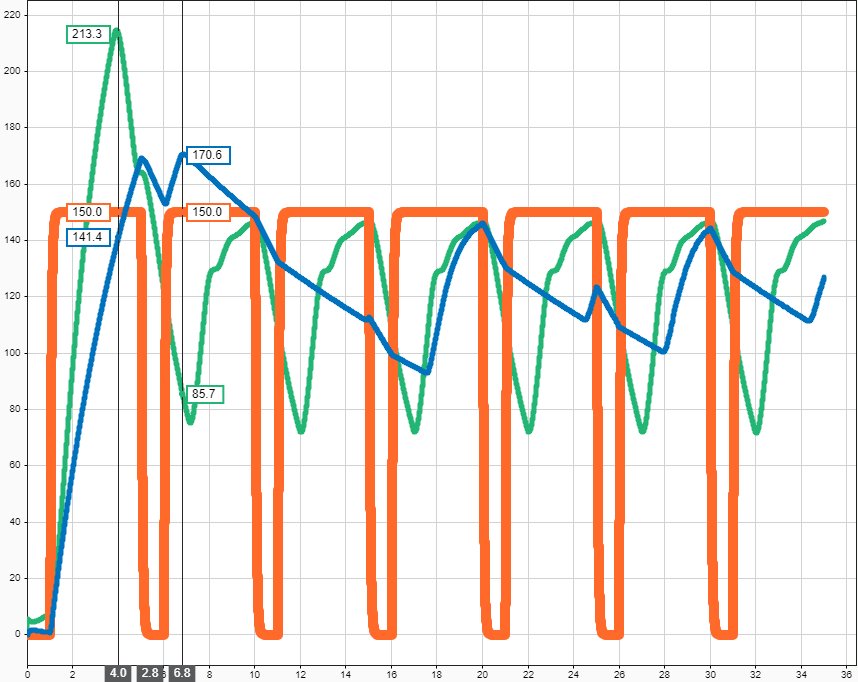}};
            \begin{scope}[x={(img.south east)},y={(img.north west)}]
                % Etichette degli assi
                \node[anchor=north, font=\scriptsize] at (0.5, -0.02) {Time (s)};
                \node[anchor=center, rotate=90, font=\scriptsize] at (-0.02, 0.5) {Speed (rpm)};
            \end{scope}
        \end{tikzpicture}
    \caption{rect-pulse-0 input.}
    \label{fig:R2_graph_step}
    \end{subfigure}
    \caption{Desired (orange) and actual speeds for the PWM (blue) and Buck (green) controllers and requirement (R2).}
    \label{fig:grafico_r2}
\end{figure*}

\Cref{fig:grafico_r2} reports the measured speed (blue) and the desired speed (orange) for two failure-revealing test cases generated by the \emph{t-pyramid-0} and  \emph{rect-pulse-0} Parameterized Test Sequences.
Specifically, for the \emph{t-pyramid-0} and \emph{rect-pulse-0} test cases, the highest values of the desired speed (orange) were respectively 167 and 150, as selected by HECATE for the \textbf{Hecate\_sp} parameter. Notice that the lower bound in each signal was 0 since these Parameterized Test Sequences corresponded to the one from \Cref{fig:parameterizedTestSequence} \emph{t-pyramid-0} and \emph{rect-pulse-0}) and no increment on the speed values was applied.

%\claudio{@Marcello puoi aggiungere un commento/riflessione personale?}

\emph{Buck Controller.} HECATE found failure-revealing test cases for all input signals with the Buck controller.
However, unlike the PWM controller, the Buck controller showed a different behavior depending on the input type.
For rectangular pulse inputs, the Buck controller showed a 10/10 falsification rate, with generally lower average ($\hat{S}$) and the median ($\bar{S}$) number of iterations required to identify the failure-revealing test case w.r.t. the PWM controller (except for the \emph{rect-pulse-130} Test Sequence).
However, HECATE did not find a failure-revealing test case for truncated pyramid Test Sequences in some of the runs with the Test Sequences encoding low-speed scenarios. Specifically, for the \emph{t-pyramid-0} and \emph{t-pyramid-85} scenarios the falsification rate was 4/10 and 5/10.
By contrast, when the Test Sequences encoded a high-speed scenario (\emph{t-pyramid-130}), HECATE returned a failure-revealing test case in all its runs (10/10 falsification rate).
%The same conclusions can be drawn from analyzing the average ($\hat{S}$) and the median ($\bar{S}$) number of iterations required to identify the failure-revealing test cases.
%In low-speed scenarios, the two values are higher (28.8 and 17.6 for the average, and 26.0 and 22.0 for the median) than in the high-speed scenario.
%This shows that is not only easier to falsify the property with Buck controller with higher speeds, but also it happens in a shorter time.

\Cref{fig:grafico_r2} reports the measured speed (blue or green) and the desired speed (orange) of two failure-revealing test cases generated by the \emph{t-pyramid-0} and  \emph{rect-pulse-0} Parameterized Test Sequences.
For the truncated pyramid Test Sequences (\Cref{fig:R2_graph_pyramid}), the behavior of the PWM and the Buck were similar, with the Buck remaining generally closer to the desired speed. 
For the rectangular pulse Test Sequences (\Cref{fig:R2_graph_step}), the Buck controller surpassed the required speed more significantly and earlier than the PWM controller.

\textbf{Safety Requirement R3.} 
We discuss the results for the safety requirement R3.
%The third requirement requires that the measured speed does not exceed the desired input speed of the e-Bike (\hecateParam{Hecate\_speed}) plus a tolerance of +1 rpm during two specific, non-continuous segments of the test scenarios: from $t=6$ to $t=10$ seconds and $t=26$ to $t=30$ seconds.

\emph{PWM controller}. 
HECATE showed a 10/10 falsification rate for each experiment for all the considered Test Sequences.  This result indicates that the PWM controller failed to maintain the speed within the desired limits in all test cases.

The average ($\hat{S}$) and the median ($\bar{S}$) number of iterations show that across all Test Sequences, HECATE required  1.0 iterations in most of the cases to identify the failure-revealing test cases both for the truncated-pyramid (\emph{t-pyramid}) and the square wave signal (\emph{rect-pulse}) Test Sequences. %The only results that are a little more than 1.0 are the ones regarding the lowest speed-value (\emph{t-pyramid-0}, \emph{rect-pulse-0}). 

% \Cref{fig:grafico_r3} reports the measured speed (blue and green) and the desired speed (orange) of two failure-revealing test cases generated by the \emph{t-pyramid-85} and  \emph{rect-pulse-85} Parameterized Test Sequences.
% For the \emph{t-pyramid-85} and \emph{rect-pulse-85} test cases, the values of the desired speed (orange) are respectively 20 and 110 since this is the value selected for the \textbf{Hecate\_sp} parameter in the analyzed segments. 
% Notice that the lower bound is 0 since these Parameterized Test Sequences correspond to the one from \Cref{fig:parameterizedTestSequence}: No increment on the speed values was applied.\claudio{Non mi torna. Non e' 85? - \emph{t-pyramid-85}}

% \claudio{@Marcello puoi aggiungere un commento/riflessione personale?}

 \emph{Buck Controller.} 
HECATE showed a 10/10 falsification rate for each experiment for all the considered Test Sequences.  This result indicates that the Buck controller failed to maintain the speed within the desired limits in all test cases.

The average ($\hat{S}$) and the median ($\bar{S}$) number of iterations show that Test Sequences focusing on truncated-pyramid Test Sequences (\emph{t-pyramid-0}, \emph{t-pyramid-85}, \emph{t-pyramid-130}) required more iterations (8.1 vs 7.3, 4.3 vs 1.6 and 2.1 vs 1.0 for the average, and 4.0 vs 1.5 and 1.5 vs 1.0 for the median) to identify the failure-revealing test cases than the ones focusing on square wave signals (\emph{rect-pulse-0}, \emph{rect-pulse-85}, \emph{rect-pulse-130}). The only result that shows an opposite trend is the median at higher speeds (\emph{t-pyramid-130}, \emph{rect-pulse-130}).

\begin{figure*}
    \centering
     \begin{subfigure}[b]{0.48\textwidth}
     \begin{tikzpicture}
            \node[anchor=south west, inner sep=0] (img) at (0,0) 
            {\includegraphics[width=.87\linewidth]{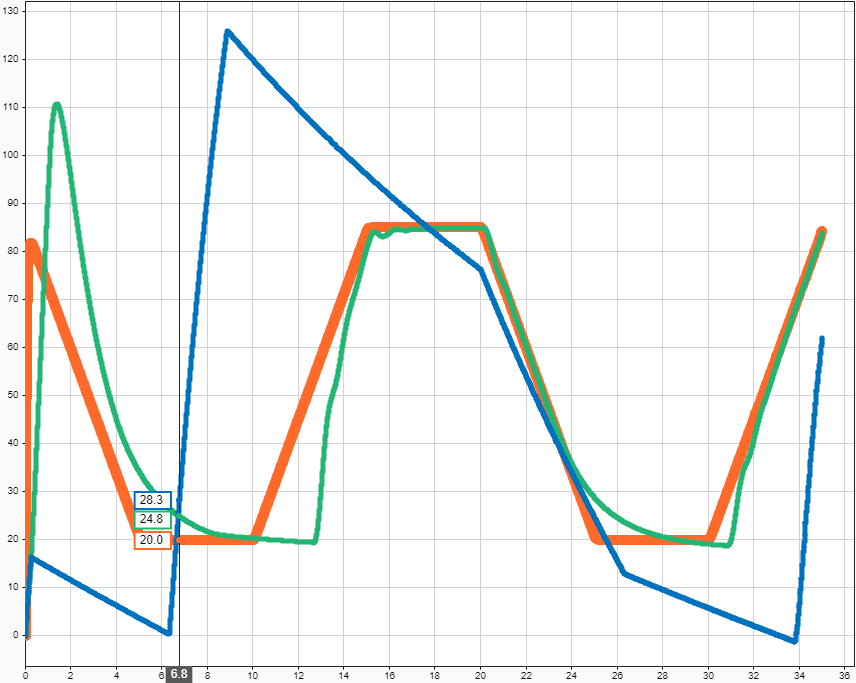}};
            \begin{scope}[x={(img.south east)},y={(img.north west)}]
                % Etichette degli assi
                \node[anchor=north, font=\scriptsize] at (0.5, -0.02) {Time (s)};
                \node[anchor=center, rotate=90, font=\scriptsize] at (-0.02, 0.5) {Speed (rpm)};
            \end{scope}
        \end{tikzpicture}
    \caption{t-pyramid-85 input.}
    \label{fig:R2_graph_pyramid85}
    \end{subfigure}
    \hfill
    \begin{subfigure}[b]{0.48\textwidth}
    \begin{tikzpicture}
            \node[anchor=south west, inner sep=0] (img) at (0,0) 
            {\includegraphics[width=.87\linewidth]{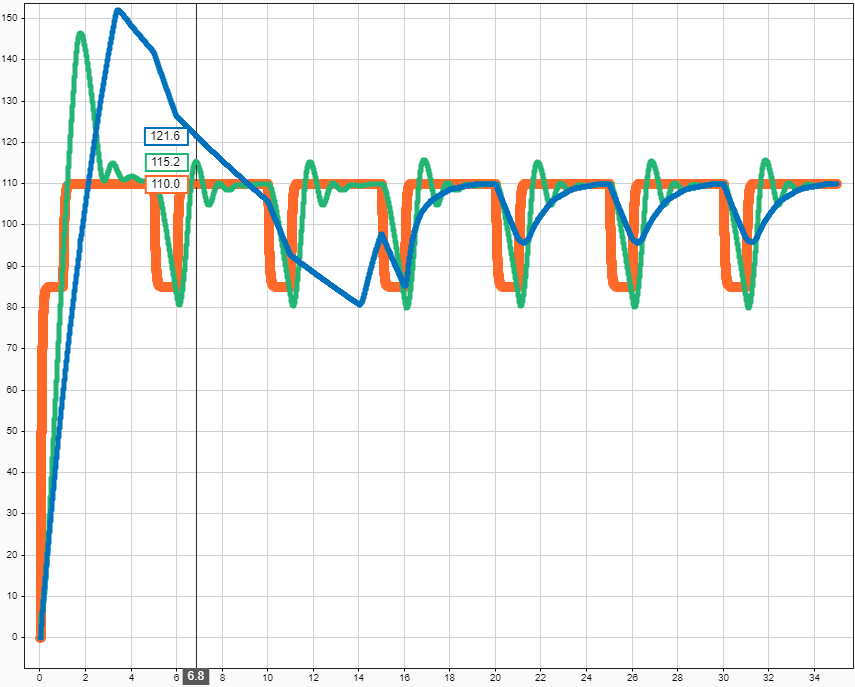}};
            \begin{scope}[x={(img.south east)},y={(img.north west)}]
                % Etichette degli assi
                \node[anchor=north, font=\scriptsize] at (0.5, -0.02) {Time (s)};
                \node[anchor=center, rotate=90, font=\scriptsize] at (-0.02, 0.5) {Speed (rpm)};
            \end{scope}
        \end{tikzpicture}
    \caption{rect-pulse-85 input.}
    \label{fig:R2_graph_step85}
    \end{subfigure}
    \caption{Desired (orange) and actual speeds for the PWM (blue) and Buck (green) controllers and requirement (R3).}
    \label{fig:grafico_r3}
\end{figure*}
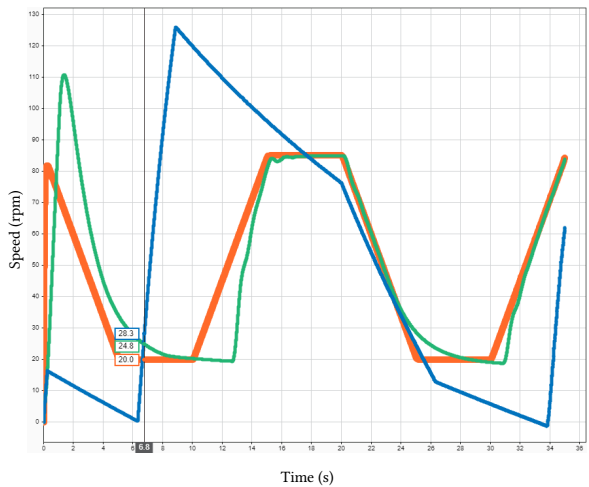
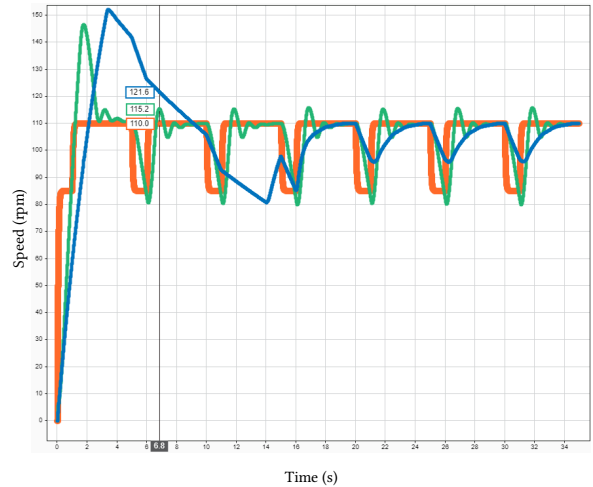

\Cref{fig:grafico_r3} reports the measured speed (blue or green) and the desired speed (orange) of two failure-revealing test cases generated by the \emph{t-pyramid-85} and  \emph{rect-pulse-85} Parameterized Test Sequences.
For the \emph{t-pyramid-85} and \emph{rect-pulse-85} test cases, the values of the desired speed (orange) were respectively 20 and 110, as selected for the \textbf{Hecate\_sp} parameter in these segments.

\textbf{Expert feedback}. 
The engineer who developed the controller models analyzed the results of our experiments and confirmed the faults we discovered.

For requirement R1, the engineer confirmed that the response (\Cref{fig:Grafico_R1_pyramid,fig:Grafico_R1_step}) to the reference signal is unstable. 
The expert hypothesis is that in some conditions, the controller continuously switches between the two algorithms (i.e., for motoring and braking functions), losing stability and reaching negative speed. 

For requirement R2, the engineer confirmed the limitations in the tracking speed during the acceleration phases and braking phases (\Cref{fig:R2_graph_pyramid} and \Cref{fig:R2_graph_step}).
For the acceleration phase and the PWM, the problem was caused by the vehicle inertia compared to the motor's power and torque. 
Note that this version of the model did not consider the cyclist torque and the bicycle gears. 
%These factors will help to have faster speed tracking in acceleration traits.
The Buck controller (\Cref{fig:R2_graph_step}) reached high speed in the first step response due to some issues in the PI (Proportional Integral) controller in the speed loop.
A more fine-grained parameter tuning could help fix this problem.
During deceleration, the braking force was not sufficient to track the reference speed. 
The engineer confirmed that the e-Bike mechanical brakes were not considered in the models.
They will be added in future versions to increase the braking force when the regenerating braking is not enough.

For requirement R3, the expert confirmed that the PWM scheme speed tracking was not accurate due to the instability caused by switching between the two different algorithms for motoring and braking functions.
Additionally, the parameters of the PI should be fine-tuned to reach a faster and more precise response.

We remark that our model is the high-level design of the e-Bike developed to compare the PWM and the Buck controllers and select the most appropriate for the considered problem.
At this development stage, the engineer was interested in the electrical variables (and not in the mechanical ones).
Therefore, some problems were expected.
The engineer also confirmed that the Buck controller presents a more advanced development stage than the PWM.

\textbf{Summary}. 
HECATE could reveal a failure-revealing test case for all 18 experiments related to the PWM controller.
HECATE could also reveal a failure-revealing test case for the 12  experiments related to the Buck controller for the regulatory and safety requirements.
Unlike the PWM controller, HECATE could not generate any failure-revealing test case for the six experiments of the Buck controller and the functional requirement.
The engineer who developed the model confirmed the test cases as failure-revealing.
These findings confirm the effectiveness of HECATE.

\begin{Answer}[Effectiveness --- RQ1]
Our results show that HECATE effectively generated failure-revealing test cases for  \percentageExperimentsFalsified (\numExperimentsFalsified out of \numExperiments) of our experiments.
\end{Answer}

\subsection{Efficiency (RQ2)}
\label{sec:efficiency}

\begin{figure}
    \centering
    \includegraphics[width=.88\columnwidth]{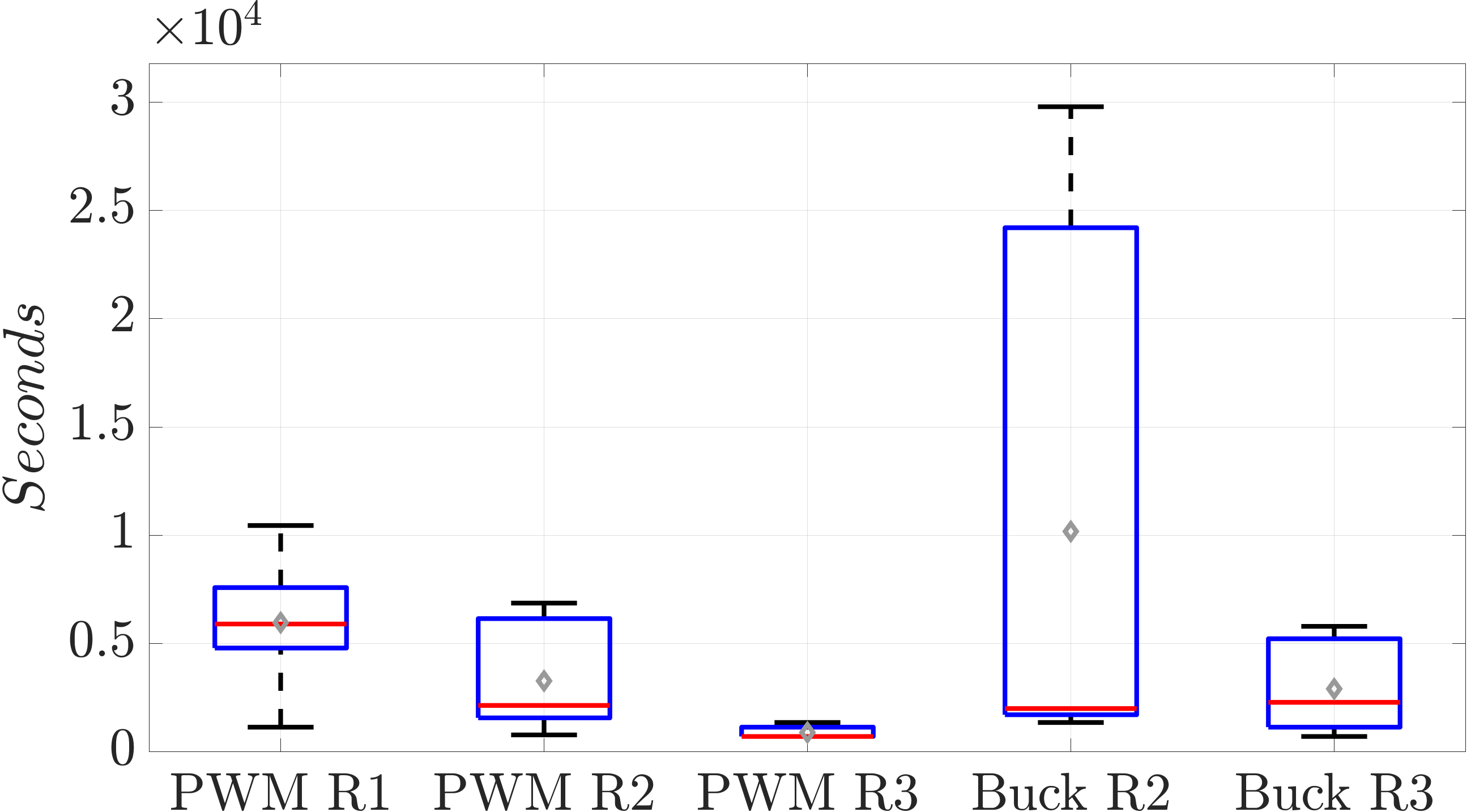}
    \caption{Time required to generate the failure-revealing test cases for all the model-requirements combinations.}
    \label{fig:boxplot}
\end{figure}

To answer RQ2, we evaluated the efficiency of HECATE in generating failure-revealing test cases for each version of the model and requirement from \Cref{sec:system} by analyzing the time required to detect a failure-revealing test case.
The boxplot from \Cref{fig:boxplot} reports our results.
This result does not include the Buck controller and the functional requirement R1, since HECATE could never produce any failure-revealing test case.  
Diamonds depict the average, and red lines represent the
median.
HECATE required, on average, $1\,min\,12\,s$ to complete a simulation ($min=1\,min\,7\,s$, $max=1\,min\,20\,s$, $StdDev=5\,s$). No significant differences in simulation times were observed between the PWM and Buck models. 

For the PWM controller, HECATE required, on average, $1\,h\,39\,min$ $26\,s$ ($min=19\,min\,5\,s$, $max=2\,h\,54\,min\,12\,s$, $StdDev=51\,min\,32\,s$) for R1, $54\,min\,41\,s$ ($min=13\,min\,8\,s$, $max=1\,h\,54\,min\,33\,s$, $StdDev=42\,min\,44\,s$) for R2, and $14\,min\,55\,s$ ($min=11\,min\,56\,s$, $max=22\,min$ $40\,s$, $StdDev=4\,min\,45\,s$) for R3.
For the Buck controller, it required, on average, $2\,h\,49\,min\,38\,s$ ($min=22\,min\,40\,s$, $max=8\,h\,16\,min\,22\,s$, $StdDev=3\,h\,39\,min\,4\,s$) for R2, and 48m31s ($min=11\,min\,56\,s$, $max=1\,h\,36\,min\,39\,s$, $StdDev=36\,min\,15\,s$) for R3. 

To compute the failure-revealing tests cases, HECATE required, on average, \averageTime (\textit{min}=\minTime, \textit{max}=\maxTime, \textit{StdDev}=\stdTime) across all failure-revealing runs of our experiments.

%Preliminary analysis on \Cref{fig:boxplot} suggests that requirements for the model with the PWM controller are more susceptible to falsification compared to those for the model with the Buck controller. Indeed, the average falsification time is $0.34 \cdot 10 ^4$ s for the PWM-based controller and $0.65 \cdot 10^4$ for the Buck-based one.

%\Cref{tab:results_time} shows the test duration for the requirements related to the analyzed reference model. Each row contains six values since each model was tested for each requirement with six different input scenarios.

%\red{TODO}

%\begin{table*}[t]
%\caption{Experimental duration results for the e-Bike case study.}
%\label{tab:results_time}
%\begin{tabular}{@{} l  |l|llllll} 
%\toprule
%\textbf{Model}  &\textbf{REQ}& \multicolumn{6}{c}{\textbf{DATA}}\\ \midrule
%\multirow{3}{*}{\textbf{PWM}}&\textbf{TA\_R1}& 6371.60& 10452,29&        4796,60& 1145,46& 5440,92&7588,65\\
% &\textbf{TA\_R2}& 6156.83 & 2147,73&       1575,00& 6872,74& 2147,73&787,50\\
% &\textbf{TA\_R3}&       1145.46 &       715,91&       715,91& 1360,23& 715,91&715,91\\ \midrule
%\multirow{3}{*}{\textbf{Buck}}&\textbf{TA\_R1}&       35795,51&       35795,51&       35795,51& 35795,51& 35795,51&35795,51\\
% &\textbf{TA\_R2}&       2076,14&       1718,18&       1932,96& 29781,86& 24197,77&1360,23\\
% &\textbf{TA\_R3}&       5226,14 &       1145,46 &       715,91 & 5798,87& 3078,41&1503,41\\ \bottomrule
%\end{tabular}
%\end{table*}

\begin{Answer}[Efficiency --- RQ2]
HECATE required, on average,  \averageTime  (\textit{min}=\minTime, \textit{max}= \maxTime, \textit{std}=\stdTime) to compute the failure-revealing test cases of our e-Bike model. 
\end{Answer}

\section{Discussion}
\label{sec:discussion}

We present lessons learned (\Cref{sec:lessons}) and threats to the validity of the findings (\Cref{sec:threats}).

\subsection{Lessons Learned}
\label{sec:lessons}
The three main lessons (L) from this study are as follows:

\textbf{L1 (Modeling)}. The engineer typically develop the tests for assessing their models by manually define inputs and visually inspect the models' outputs.
This is often a common practice in industrial environments especially when preliminary and high-level models of the system (like the one we considered) are evaluated.

The engineer who developed the two models confirmed that the proposed Test Sequence blocks helped reflect on plausible inputs and their characteristics, and Test Assessment blocks helped formalize the requirements of their system.
Therefore, the engineer confirmed the usefulness of the rigorous formalization of the test inputs and the system requirements.

\textbf{L2 (Testing Procedure)}. The outputs provided by the proposed testing framework helped the engineer identify flaws within the system design (\Cref{sec:effectiveness}). 
The time required to compute the failure-revealing test cases was practical for industrial applications (\Cref{sec:efficiency}).

\textbf{L3 (Comparison of Solutions)}. HECATE (and SBST in general) was beneficial for assessing the benefits and limitations of different controller implementations.
Specifically, in our case, based on the feedback provided by HECATE, we were able to assess when the PWM and Buck worked properly and select in which scenario using each controller can be beneficial.

\begin{figure}
    \centering
    \includegraphics[width=\columnwidth]{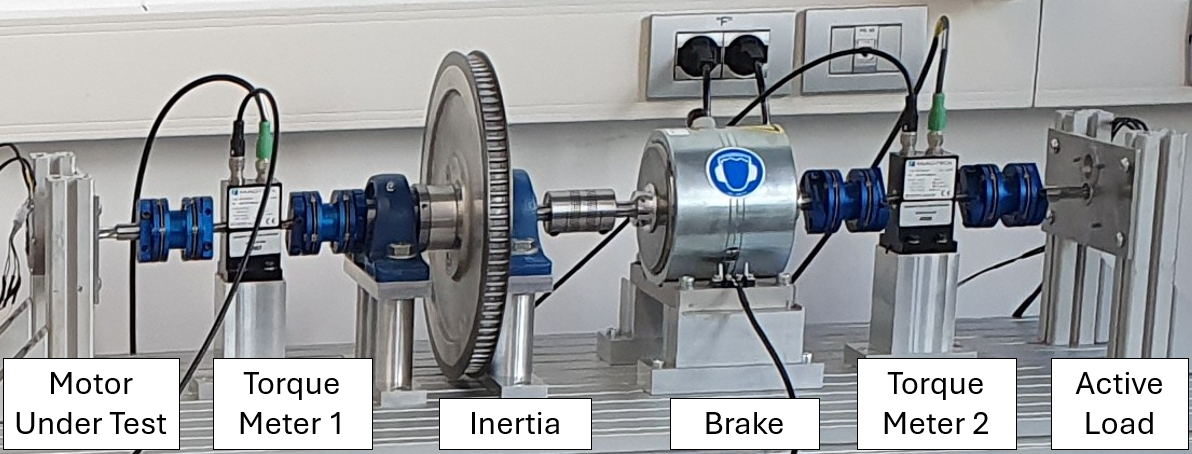}
    \caption{Physical test bench.}
    \label{fig:testBench}
\end{figure}

Considering these lessons, the engineer is now using SBST and HECATE to automatize their testing procedure. 
The engineer has also indicated their interest in evaluating the failure-revealing test case on the physical platform reported in \Cref{fig:testBench}, which they are currently finalizing. 
The physical platform consists of the Motor Under Test, a torque meter, a modulable inertia disk, a magnetic hysteresis brake, a second torque meter, and an active load.

Our results are relevant for \emph{industrial application} of testing procedures. 
Technology transfer activities require empirical studies that industries can use to assess the effectiveness of different technologies.
Our results show the effectiveness of HECATE in detecting failure-revealing test cases for a complex e-Bike powertrain model developed within a project involving industrial partners.

Our results significantly \emph{improved the state of practice} of testing procedures. 
Prior to our study, engineers typically developed their models before manually developing their test 
cases.
%Since the model was in a preliminary development stage, they were testing its behavior by considering profiles for smooth limited changes in the desired speed. 
During the preliminary development stage of the proposed models, engineers tested their behavior by considering profiles for smooth 
limited changes in the desired speed. 
This project improved the state of practice by showing the benefits and effectiveness of the search-based approach implemented by HECATE.
The data related to the effectiveness of SBST can also benefit other practitioners developing similar CPSs that are considering the SBST technology.
%Questo lo conferma
%Industrial

%Industrial application: The applied research or experience report has taken place in an industrial setting and/or with an industrial partner. Research on a problem relevant to industry or motivated by current and forthcoming industrial challenges is also welcome.

%loro non stavano facendo testing ah ah gli e' piaciuto molto.
%Improvement on the state of the practice: The amount of improvement that the work achieves above and beyond the state-of-the-practice.
%Clarity of lessons learned: The clarity in which the lessons learned are presented and how well they are supported with data and discussion.

%There are two main \emph{lessons learned} from our experiments.
%First, the engineers developing the e-Bike model learned and confirmed the benefits of using HECATE during the development of their models. 
%This lesson is confirmed by the reflections from the expert about the failures reported in \Cref{sec:evaluation}.
%Second, from the testing perspective, we learned that HECATE (and SBST in general) can also be beneficial for assessing the benefits and limitations of different implementations.
%For example, to assess when PWM and Buck work properly and select when the use of each controller is beneficial.

% stabilire pro e control di implementazioni differenti

%Generality of results: A clear discussion about how the work, approach, or lessons learned are applicable to practitioners outside of the studied group.
Concerning the \emph{generalizability of our results}, we do not expect that applying HECATE to other systems will return the same percentage of failure-revealing test cases.
This percentage strongly depends on the model, its development stage, and the Test Blocks selected for running HECATE.
However, our results are \emph{general}: they confirm existing results from the research literature obtained in other domains (e.g., space~\cite{Menghi2020}, automotive~\cite{fainekos2012verification,stocco2023model}, 
biomedical~\cite{ayesh2022two}, 
medical~\cite{arcaini2016model,ayesh2022two,Bombarda2022}).
Furthermore, our results confirm that the previous results reported on HECATE~\cite{Formica2023,Formica2024} are also applicable to the e-Bike domain.
%Additional empirical studies will confirm whether these results will be generalized to CPS from other domains.

%In terms of general performance, both models successfully completed all 18 tests, showing varied levels of performance across different scenarios. These variations underscored the distinct characteristics and capabilities inherent in each model's design. A key highlight to the results revealed that the models responded differently to the Test Assessment scenarios, with Buck model generally meeting more of the performance requirements outlined in  \Cref{tab:reqs} compared to PWM model. Notable observations emerged from specific scenarios, where unexpected discrepancies were detected between the measured speed and the intended input profile. A more detailed analysis of these findings will be presented in the next section to further explore the reasons behind these discrepancies and their implications. 

\subsection{Threats to Validity}
\label{sec:threats}
The requirements and the Parameterized Test Sequences we considered in this study could threaten the \emph{external validity} of our results. 
However, the fact that the requirements and the Test Sequences were defined in collaboration with the engineer who developed the model mitigates this threat.
The selection of our study subject (a model from the e-Bike domain) could threaten the \emph{external validity} of our results. 
We do not claim that our results can be generalized to study subjects from other domains. 
However, the fact that our study subject is a representative model developed by expert engineers within a project involving industrial partners mitigates this threat. 
Our results confirm the findings from the research literature~\cite{Formica2023}.
Therefore, they are likely generalizable to other systems.
%Other industry papers will provide additional empirical evidence, or reject our hypothesis for other models and systems.
Future industrial studies are needed to provide additional empirical evidence or refute our hypothesis in other models and systems 

%\andrea{@Claudio: Il paragrafo seguente è da riscrivere? Sembra manchi una frase alla fine}
%\claudio{Yes, da riscriere e' preso dall'altro paper %thank you.}
The values assigned to the configuration parameters selected for HECATE could
threaten the \emph{internal validity} of our results.
For example, considering more iterations for our SBST framework or a different search algorithm can lead to different results. 
To mitigate this threat, we reused the default values for the configuration parameter provided by HECATE. 

%\claudio{Da inserire: abbiamo usato scenari proposti dall'ingegnere come ispirazione (mitigazione)}

\section{Related Work}
\label{sec:related}
Numerous studies have evaluated the effectiveness of SBST in identifying failure-revealing test cases for CPS development~\cite{Ahmed2020,Bombarda2022,Formica2023,Ling2023,Formica2023}.
In this work, we assessed the usefulness of SBST by considering the motor controller for an e-Bike, analyzing two different implementations, namely one based on a Buck converter and one controlled by using the PWM strategy.

Testing e-Bike motor controllers is of utmost importance, especially given the ever-increasing complexity of these vehicles.
However, this activity is commonly performed by physically testing electric bikes or their components with different loads, pedaling profiles, roads and scenarios~\cite{Parastiwi2020, Abagnale2016, Gupta2024, Lefticaru2017}.
Instead, in this work, we used HECATE~\cite{Formica2024} for model-in-the-loop testing.

\section{Conclusion}
\label{sec:conclusion}

This industrial paper presents our assessment of the effectiveness and efficiency of HECATE in generating failure-revealing test cases for an e-Bike system. 
HECATE successfully identified failure-revealing test cases in practical time. 
The failure-revealing test cases were confirmed by the engineer who developed the model. 
We critically reflected on our results, presented lessons learned, and discussed the relevance of our results for industrial applications. 
Finally, we discussed how our findings improved the state of practice and the generalizability of our results.

\section*{Data Availability}
A replication package containing all of our data, test results, and scripts is publicly available~\cite{ExperimentalResults}. 

\begin{acks}
This work was supported in part by project SERICS (PE00000014) under the NRRP MUR program funded by the EU - NGEU, by the European Union - Next Generation EU. ``Sustainable Mobility Center (Centro Nazionale per la Mobilità Sostenibile - CNMS),'' M4C2 - Investment 1.4, Project Code CN\_00000023, and by PNC - ANTHEM (AdvaNced Technologies for Human-centrEd Medicine) - Grant PNC0000003 – CUP: B53C22006700001 - Spoke 1 - Pilot 1.4.
\end{acks}

\bibliographystyle{ACM-Reference-Format}
\bibliography{mybibliography}

%\clearpage
%\section*{Appendix A: Demonstration walkthrough}
%\input{walkthrough}
\clearpage

%\section*{Appendix B: Installation and use}
%\input{use}
%\input{installation}

\end{document}